\documentclass[11pt]{article}

\usepackage[margin=1in]{geometry}
\usepackage[T1]{fontenc}
\usepackage[utf8]{inputenc}
\usepackage{graphicx}
\usepackage{url}
\usepackage{float}
\usepackage{amsmath}
\usepackage{amssymb}
\usepackage{latexsym}
\usepackage{bm}
\usepackage{array}
\usepackage{tabularx}
\usepackage{booktabs}
\usepackage[round,authoryear]{natbib}
\usepackage[hidelinks]{hyperref}

\newcommand{\correspondingemail}{ebouzeid@princeton.edu}

\begin{document}

\title{Data-Driven Flux Parameterization for the Atmospheric Boundary Layer}

\author{Abed Hammoud\textsuperscript{1}, Edriss S. Titi\textsuperscript{2}, Mitchell Bushuk\textsuperscript{3}, Marc Calaf\textsuperscript{4},\\
Khaled Ghannam\textsuperscript{5}, and Elie Bou-Zeid\textsuperscript{1}\\[0.75em]
\small \textsuperscript{1}Civil and Environmental Engineering, Princeton University,\\
\small Princeton, New Jersey 08540, USA\\
\small \textsuperscript{2}Department of Applied Mathematics and Theoretical Physics, University of Cambridge,\\
\small Cambridge CB3 0WA, UK\\
\small \textsuperscript{3}National Oceanic and Atmospheric Administration/Geophysical Fluid Dynamics Laboratory,\\
\small Princeton, New Jersey 08540, USA\\
\small \textsuperscript{4}Department of Mechanical Engineering, University of Utah,\\
\small Salt Lake City, Utah 84112, USA\\
\small \textsuperscript{5}Department of Civil and Environmental Engineering, Northeastern University,\\
\small Boston, Massachusetts 02115, USA\\[0.75em]
\small Corresponding author: Elie Bou-Zeid (\texttt{\correspondingemail})}

\date{}
\maketitle

\section*{Key Points}
\begin{itemize}
\item Linear operators learn turbulent fluxes from boundary layer profiles
\item Stable cases show local mixing while convective cases show nonlocal transport
\item Single column tests stay numerically stable across abrupt regime transitions
\end{itemize}

\begin{abstract}
Turbulent fluxes in the atmospheric boundary layer (ABL) govern the exchange of momentum, heat, and mass between the surface and the atmosphere, thereby shaping the vertical structure of the ABL and influencing a wide range of engineering applications, atmospheric processes, and boundary-layer dynamics.
Accurate parameterization of these fluxes in coarse-resolution weather and climate models remains challenging, particularly under unstable conditions where flux transport is strongly nonlocal and under stable conditions where turbulence may become intermittent and partially decoupled from surface forcing.
In this study, we develop a data-driven parameterization of turbulent fluxes using a novel formulation in which nondimensional fluxes are represented as a linearized convolution operation acting on the relevant nondimensional mean-state profiles.
A comprehensive high-resolution large-eddy simulation (LES) dataset is generated for idealized flow over homogeneous surfaces across a range of stability conditions.
Using selected training cases from this dataset, we derive and test several first-order versions of this linearized convolution flux parameterization, each based on different combinations of inputs, that map mean temperature and velocity profiles to heat and momentum fluxes.
The best-performing parameterization is determined as the one that minimizes the mean squared error in both training cases and unseen testing cases.
The results indicate that the proposed method improves predictive skill relative to a standard K-profile parameterization (KPP) benchmark while preserving a transparent operator structure that can be inspected for physical meaning.
Rather than treating the learned closure as a purely empirical mapping, the present framework learns flux--profile relationships as operators constrained by the governing column dynamics, enabling the inferred operator kernels to reveal the spatial organization, locality, and nonlocality of turbulent transport across stability regimes.
Despite being restricted to linear operators, the resulting parameterization remains interpretable, performs strongly across diverse stability conditions, and tests successfully in \textit{a posteriori} single-column model simulations, yielding state profiles that agree closely with the LES data.
These findings suggest that data-informed operator closures can relax restrictive assumptions embedded in traditional closure schemes while providing a useful language for interpreting turbulent transport, guiding practical implementation in column models, and informing future theoretical developments for ABL parameterization.
\end{abstract}

\section*{Plain Language Summary}
Accurate representation of turbulent fluxes in the atmospheric boundary layer is critical for various engineering applications and for reliable weather and climate prediction, yet parameterizations remain particularly challenging due to the various complex physical processes involved. 
This study introduces a data-driven framework, based on a new mathematical formulation of the problem, where datasets from turbulence-resolving simulations are used for deriving physically interpretable flux parameterization schemes. 
By formulating the fluxes as a convolution operation between a learnable kernel with mean profiles, we capture both nonlocal effects in convective regimes and the local scaling in stable regimes. 
This framework reduces reliance on idealized assumptions and provides a flexible pathway toward physically grounded, data-informed turbulent flux models.

\section{Introduction}
\label{sec:intro}

Turbulent fluxes in the atmospheric boundary layer (ABL) govern the vertical transport of momentum, heat, moisture, and mass, thereby shaping the coupling between the surface and the free troposphere \cite{stull2012introduction}. 
These fluxes shape wind and temperature profiles, regulate entrainment and ABL growth, and redistribute moisture--processes that govern cloud development, surface energy exchange, and ultimately, the predictability of weather and climate.
However, the turbulent eddies and convective plumes that dominate vertical transport occur at scales smaller than those resolved by current general circulation models (GCMs) or numerical weather prediction (NWP) systems. 
Despite steady increases in computational power, explicitly resolving ABL turbulence and shallow convection remains computationally prohibitive in GCMs and introduces ``gray zone'' challenges in finer resolution NWPs where turbulence is partially resolved \cite{Wyngaard2004, Tomassini2023}.
Consequently, parameterizations will remain required to represent the aggregate effects of unresolved turbulence on resolved model variables for the foreseeable future \cite{Suselj2012, HanBretherton2019, LopezGomez2020}.

Conventional parameterization approaches rely on gradient-diffusion closure, in which fluxes are related to mean gradients through an eddy diffusivity coefficient derived from mixing-length concepts \cite{Prandtl1925, Taylor1915, Blackadar1962}. 
While early closures were purely local, analogous to molecular diffusion models, subsequent work introduced nonlocal contributions to capture mixing produced by large coherent eddies and entrainment processes. 
The K-Profile Parameterization (KPP) provides a widely used formulation of this class, initially outlined for the ABL by \citet{TroenMahrt1986}, extended by \citet{HoltslagBoville1993}, and later adapted for oceanic boundary layers \cite{Large1994}. 
KPP prescribes a vertical diffusivity profile based on surface-layer similarity scaling, a boundary layer shape function, and entrainment matching near the ABL top, with model extensions introducing a nonlocal transport term to represent large-eddy effects \cite{lock2000new}. 
% This framework has been incorporated into several GCMs and NWP systems due to its computational efficiency and ability to unify local and nonlocal transport. 
% Nevertheless, 
Although computationally efficient, KPP relies on empirically tuned coefficients and rigid assumptions regarding mixing lengths and entrainment, which limit its universality across diverse stability regimes and motivate efforts toward scale-aware and data-driven alternatives \cite{Garanaik2023, Yuan2024}.

The eddy-diffusivity mass-flux (EDMF) framework was subsequently introduced to overcome some of these shortcomings by unifying local diffusive mixing and nonlocal convective transport \cite{Siebesma2007, Soares2004}. 
EDMF represents unresolved fluxes as the sum of an eddy-diffusivity term, capturing small-scale turbulence, and a mass-flux term, representing larger, but still unresolved, coherent thermal plumes. 
This dual formulation provides a consistent treatment of turbulence across dry convective, cloud-topped, and stable boundary layers, and has been widely adopted in weather and climate models \cite{Suselj2012, HanBretherton2019, Perrot2025, Tan2026}. 
Recent developments have incorporated prognostic turbulence kinetic energy (TKE) closures and stochastic multi-plume representations, validated through large-eddy simulations (LES) \cite{LopezGomez2020, Suselj2019Factors}. 
However, despite these advances, EDMF and related parameterizations remain constrained by structural assumptions and empirical closures that may not generalize across ABL flow regimes \cite{Tan2018, Cohen2020, Strobach2022, ghannam2026coupling}.

The limitations of fixed-form, empirically tuned closures, together with the growing availability of high-resolution simulations and observations, have motivated data-driven and hybrid parameterizations for climate and weather models. 
In the ocean, early work demonstrated that machine learning can infer mesoscale eddy closures from coarse-grained high-resolution fields. 
For instance, \citet{Bolton2019} used deep-learning methods for ocean data inference and subgrid parameterization, while \citet{Zanna2020} used equation-discovery and convolutional approaches to obtain interpretable mesoscale closures. 
Subsequent studies moved these ideas toward online use in ocean general circulation models, including stochastic neural parameterizations of subgrid momentum forcing \cite{Guillaumin2021}, implementations of learned mesoscale forcing in MOM6 \cite{Zhang2023}, and stable, scale-aware or physics-constrained forms designed to improve numerical robustness and generalization across resolutions \cite{Perezhogin2024, Perezhogin2025}. 
In the atmosphere, early neural network parameterizations for convection and cloud-resolving physics \cite{Krasnopolsky2013, Gentine2018, OGorman2018, Rasp2018, Brenowitz2018} established the feasibility of replacing or augmenting conventional closures, while later studies emphasized online stability, physical constraints, and generalization \cite{Yuval2020, Yuval2021, Beucler2021, Wang2022, WattMeyer2024}.
A complementary direction embeds learned components within process-based closure frameworks, especially EDMF schemes for turbulence and convection \cite{Schneider2017, Cohen2020, LopezGomez2020}, which enabled gradient-free ensemble Kalman calibration of EDMF parameters \cite{LopezGomez2022} and online learning of entrainment closures in hybrid EDMF schemes \cite{Christopoulos2024}.

Most recently, \citet{zanna2025frameworkhybridphysicsaicoupled} introduced a framework for hybrid physics–AI climate models that integrates machine-learned parameterizations of ocean mixing and mesoscale fluxes into operational circulation models, demonstrating stable online coupling and highlighting the importance of physical constraints, scale-awareness, and training strategies for robust deployment.
Nevertheless, most previous data-driven schemes remain constrained by either by predefined functional forms and empirical tuning (e.g., data-driven EDMF) or by low interpretability (e.g., neural networks and related approaches). 
In addition, their generalizability across the full range of static stability regimes observed in the ABL is still untested, which can hinder their robustness when embedded in prognostic atmospheric models.

These challenges motivate the development of a more flexible, yet physically interpretable, framework for representing vertical turbulent fluxes.
A particularly appealing approach is operator learning, wherein solution operators of dynamical systems are approximated directly from data \cite{Hammoud2022}. 
Operator learning, hence, seeks a ``rule" that maps entire input functions (e.g., mean-state profiles) to output functions (e.g., turbulent flux profiles) \cite{BOULLE202483}.
Rooted in the theory of Green's functions and convolutions, this approach enables the development of surrogates that are resolution-independent and require minimal a priori structural assumptions. 
By treating the ABL flux at a given height as a linearized convolution of the entire vertical profile, one can explicitly account for nonlocal transport and dependence on higher-order gradients, while maintaining a level of mathematical transparency that reveals the underlying physical drivers of the flow.

In this work, we propose a data-driven parameterization of vertical turbulent fluxes in the ABL, formulated as a linearized operator learned from LES data. 
A novel mathematical formulation is developed to express vertical fluxes as a convolution operation between a kernel and the mean state. 
By linearizing this expression, the unresolved turbulent flux is approximated as the product of a linear operator, which describes the parameterization, and the resolved vertical profiles of the mean state. 
Rather than prescribing a fixed $K$-profile or entrainment law, the operator structure and magnitude are inferred directly from the high-fidelity data. 
This framework generalizes existing closures where traditional diffusive and mass-flux schemes emerge as special cases, while the learned operator captures a broader range of mixing dynamics. 
The learned operator is trained and validated against LES data of ABL flows across a range of stability regimes. 
Its performance is further evaluated in a single-column model (SCM) framework across a wide range of surface heat fluxes and cooling rates, thereby assessing its robustness and applicability under diverse boundary-layer conditions.
This linear-operator representation provides interpretability, numerical stability, and extensibility, and naturally aligns with emerging operator-learning techniques in scientific machine learning. 
Our objective is to position this framework as a physically grounded, data-informed pathway toward next-generation turbulence parameterizations that unify empirical knowledge with high-resolution data across flow regimes.

\section{Methods}
\label{sec:methods}
This section introduces the physical and data-driven flux parameterizations as well as the numerical details of the LES runs used for model training and validation.

\subsection{Flux Parameterization}
\label{ssec:math}

In ABL modeling for weather or climate simulations, the need for flux parameterization arises because turbulence transports momentum, heat, and scalars across a wide range of scales, most of which are not explicitly resolved by the coarse-scale model numerical grid.
The standard approach is to solve the Reynolds-averaged Navier-Stokes (RANS) equations for the mean state and model the turbulent fluxes.
Mathematically, this consists of decomposing the flow field into mean and fluctuating components using Reynolds decomposition: $\phi = \overline{\phi} + \phi^{\prime}$, where $\overline{\phi}$ denotes the Reynolds-averaged quantity and $\phi^{\prime}$ represents the turbulent fluctuation. 
The property $\phi$ typically represents a prognostic state variable such as potential temperature $\theta$ or any of the velocity components ($u \equiv$ streamwise, $v \equiv$ lateral, $w \equiv$ vertical) along the corresponding spatial coordinates ($x$, $y$, $z$).
Substituting such decompositions into the governing equations, under the assumption of horizontal homogeneity, and averaging leads to additional unknown terms in the equations of the mean states, such as the Reynolds shear stresses $\overline{u' w'}$ and heat flux $\overline{w'\theta'}$. 
These terms represent vertical turbulent transport, but introduce a ``closure problem'', since the number of unknowns exceeds the number of governing equations.

Flux parameterization, therefore, seeks to express these turbulent covariances in terms of the resolved mean fields.
The simplest closure form parameterizes these fluxes using downgradient diffusion (KPP closure), such that the flux is determined by the local mean gradient:
\begin{equation*}
    \overline{u'w'} = -K_m \frac{\partial \overline{U}}{\partial z}; 
\qquad 
\overline{w'\theta'} = -K_h \frac{\partial \overline{\theta}}{\partial z},
\end{equation*}

\noindent where $K_m$ and $K_h$ are the eddy diffusivities for momentum and heat, respectively. 
A practical difficulty lies in specifying these diffusivities and their dependence on stability, height, and turbulence length scales, especially under strongly stratified or convective conditions.
Alternatively, this parameterization can be reformulated and generalized for any property $\phi$ as the matrix-vector product: $\overline{w'\phi'}\approx A \overline{\phi}$, where A is a linear operator mapping the mean profile $\overline{\phi}$ to its corresponding flux.
This linear operator, in this case, represents the prescribed K-profile and mathematical operations, such as differentiation, where $A = -K_{\phi} \partial/\partial z$.
This linearized operation models the flux parameterization through the lens of a linear operator (A) that operates on, but is assumed to be independent of, the velocity or temperature profiles, though it can depend on the boundary conditions (e.g., surface fluxes).
In operational implementations of KPP, the diffusivities of heat and momentum often depend on the gradients of velocity and temperature. 
While we will show later that this is not required if the operator is formulated more generally, a linear interdependence of the flux-profile relations for temperature and velocity will be reintroduced.

A more recent flux parameterization is the EDMF scheme \cite{Soares2004, Siebesma2007,Suselj2012, ghannam2017non}, a hybrid approach developed to represent ABL fluxes as a superposition of a local contribution, modeled via downgradient diffusion, and nonlocal transport by convective updrafts, modeled via a mass flux approach. Thus, the turbulent flux in EDMF is parameterized as: 
\begin{equation*}
\overline{w'\phi'} = -K_\phi \frac{\partial \overline{\phi}}{\partial z} 
\;+\; \sum_{i=1}^{N_e} a_i \left( \phi_i - \overline{\phi} \right) w_i,
\end{equation*}
\noindent where $K_\phi$ is the eddy diffusivity; $a_i$ is the fractional area occupied by the $i$-th updraft; $\phi_i$ and $w_i$ are the scalar property (e.g., potential temperature) and vertical velocity within that updraft; and $N_e$ is the number of updrafts. 
In Appendix \ref{app:deriv}, we show that this EDMF parameterization can be formulated as:
\begin{equation}
    \overline{w'\phi'} \approx C * \overline{\phi},
    \label{eqn:conv}
\end{equation}
\noindent where `$*$' represents a convolution operation between a kernel $C$ and the mean profile $\overline{\phi}$.
By linearizing the convolution form of the equation, the parameterization also reduces to a matrix-vector product, where the eddy diffusivity model is a unique case thereof:
\begin{equation}
    \overline{w'\phi'}\approx A \overline{\phi}.
\end{equation}

The linear operator $A$ can be learned by first generating a dataset of input-output data pairs ($\overline{\phi}$, $\overline{w'\phi'}$).
Note that these data pairs were generated from LES in this study by considering time averages over $N_t$ different periods extracted from the full solution trajectory (similar, for example, to the averaging windows for observational turbulence data analysis), where the stability regime dictates the duration for which the averaging becomes suitable to learn a statistically-converged operator. 
Without loss of applicability, remark that this learned operator maps the mean profiles to their corresponding fluxes, meaning it can be applied at any time step in a RANS simulation (which is averaged by construction), regardless of the averaging period used to derive this operator, making it particularly attractive for \textit{a posteriori} testing. 
That is, it is a RANS closure that can be used in weather or climate models even when their time steps are smaller than the averaging period that would be required to converge the turbulence statistics.
Finally, we note that the operator $A$ can be derived by solving the inverse problem:
\begin{equation}
\begin{bmatrix}
| &  &  | \\
\overline{w'\theta'}_1 & \ldots & \overline{w'\theta'}_{N_t} \\
| &  & | 
\end{bmatrix}_{N_z \times N_t}
=
A \cdot 
\begin{bmatrix}
| &  &  | \\
\overline{\theta}_1 & \ldots & \overline{\theta}_{N_t} \\
| &  & | 
\end{bmatrix}_{N_z \times N_t}
\end{equation}
\noindent where the left-hand side is constructed by stacking the $N_t$ time-averaged vertical flux profiles as columns, and the right-hand side similarly with the mean profiles. 
A plethora of options are available to solve this system, and in this study, we rely on the Tikhonov regularized regression approach because it shows comparable results with alternative methods at a fraction of the computational costs \cite{Kim2007_LASSOcompute}. 
Also note here that, while these parameterizations are learned on a vertical grid of resolution $N_z$, it can be extended to larger/smaller grids by stretching/contracting the linear matrix through affine linear transformations, for example, interpolation. 
Generalizability to different resolutions is touched on in Sections \ref{ssec:scm_neutralIC} and \ref{ssec:scm_stabilityICs}.

Note that this formulation draws natural links to computational tools for learning reduced-order linear operators, such as Koopman operator theory \cite{Brunton2022}, which lifts nonlinear dynamics to a space of observables where evolution is linear, and Dynamic Mode Decomposition \cite{Schmid2010, Williams2015_OpLearn}.
These methods advance any observable by composition with the flow, and its spectral objects (eigenvalues, eigenfunctions, modes) expose coherent structures and time scales through low-rank linear surrogates.
On the other hand, the present approach learns time-independent parameterizations (quasi-stationary), where the parameterization consists of a linear operator embedding all the numerical discretization related to the physical processes (e.g.\ derivatives, non-local mixing, eddy diffusivity,...).

% More recently, nonlinear operators through neural networks have been developed to learn the effective dynamics of nonlinear systems from data. 
% This direction historically started using artificial neural networks \cite{GonzalezGarcia1998}, before being further extended to Fourier Neural Operators \cite{Li2021FNO} and DeepONets \cite{Lu2021} to name a few. 
% In fact, operator learning has shown great advancement for reliably summarizing the collective effect of nonlinear dynamics in various applications \cite{Lanthaler2025}.

The formulation can be further extended to allow for other state variables to be incorporated into the parameterization. 
For instance, the heat flux can be estimated using the mean temperature as well as the velocity components according to:
\begin{equation}
\label{eq:heat}
    \overline{w'\theta'} \approx A_{\theta}\overline{\theta} + A_u\overline{U} + A_v\overline{V},
\end{equation}
where $A_{\theta}$, $A_{u}$ and $A_{v}$ are linear operators acting on the mean temperature $\overline{\theta}$ and the mean horizontal velocity components $\overline{U}$ and $\overline{V}$. 
The physical rationale in Equation \ref{eq:heat} in the diabatic ABL is that the eddy viscosity and diffusivity depend on the mechanical and buoyant production of TKE, such that the velocity and temperature fields interact strongly. 
This resolves the aforementioned limitation, imposed to keep the model linear, where the $A$ matrices are independent of the flow variables, allowing the velocity field to influence the temperature fluxes and vice versa. 
In this study, we evaluate different input configurations and normalize the input–output pairs (as detailed later) to ensure our method generalizes across diverse physical regimes. 
For brevity, we present results for a subset of input combinations in the main text, and provide the remaining cases in the Supplementary Material.

\subsection{Numerical details of the LES algorithm}
\label{ssec:les}

To generate the high-fidelity datasets required for operator learning, this study conducts and utilizes a suite of LES experiments spanning a range of stability conditions and forcing. LES has become a well-established ``virtual laboratory" in the study of high Reynolds number ABL flows, as reviewed by \citet{stoll2020large} and \citet{moeng2015large}, where the energy-containing scales are explicitly resolved on a numerical grid, while the effects of small-scale eddies are approximated via a subgrid-scale (SGS) model \cite{Meneveau2000, BouZeid2014}. This approach provides three-dimensional, time-resolved fields of turbulent fluxes and mean profiles under controlled conditions, making it an ideal source for training and validating data-informed closures.

The work here uses an in-house LES algorithm \cite{BouZeid2004, BouZeid2005} that has been extensively validated against similarity laws and experiments for various stabilities \cite{Kumar2006,Huang2013a} and terrain morphologies \cite{Li2016}, and used to study diverse regimes including baroclinic ABLs \cite{momen2018modulation, ghannam2021baroclinicity}, and flows and thermal circulations over heterogeneous terrain  \cite{fogarty2024numerical, allouche2025unsteady}.
The code solves the spatially filtered mass, momentum, and thermal energy conservation equations for an incompressible flow with the Boussinesq approximation. 
The governing equations are:

\begin{equation}
\label{eq:LES1}
\frac{\partial \tilde{u}_i}{\partial x_i} = 0 ,
\end{equation}

\begin{equation}
\label{eq:LES2}
\frac{\partial \tilde{u}_i}{\partial t}
+ \tilde{u}_j \left(\frac{\partial \tilde{u}_i}{\partial x_j} - \frac{\partial \tilde{u}_j}{\partial x_i} \right)
= - \frac{1}{\rho_r} \frac{\partial \tilde{p}^{*}}{\partial x_i}
+ g \left(\frac{\tilde{\theta} - \theta_r}{\theta_r}\right)\delta_{i3}
- \frac{\partial \tau_{ij}}{\partial x_j}+ \tilde{F}_i,
\end{equation}

\begin{equation}
\label{eq:LES3}
\frac{\partial \tilde{\theta}}{\partial t}
+ \tilde{u}_j \frac{\partial \tilde{\theta}}{\partial x_j}
= -\frac{\partial q_j}{\partial x_j}.
\end{equation}

\noindent where the $\tilde{.}$ symbol denotes spatial filtering at scale $\Delta = \left(dx \, dy \right)^{1/2}$; $\tilde{u}_i$ ($i =$ 1, 2, 3) is the resolved/filtered velocity field in the three Cartesian directions $x_i$ (horizontal: $x$, $y$, and vertical: $z$); and $\tilde{\theta}$ is the resolved potential temperature; and $\rho_r$ is a reference Boussinesq density corresponding to the reference potential temperature defined as the average over the $x-y$ plane, $\theta_r=\langle\tilde{\theta}\rangle_{xy}$. 
The convective term in Equation \ref{eq:LES2} is written in rotational form to ensure kinetic energy conservation by the inertial terms \cite{kravchenko1997effect}, and $\tilde{p}^{*} = \tilde{p} + \frac{1}{3}\tau_{kk} + \frac{1}{2}\tilde{u}_j\tilde{u}_j$ is the dynamic modified pressure that accounts for the SGS and the resolved kinetic energy. 
The buoyancy term $ g \left(\frac{\tilde{\theta} - \theta_r}{\theta_r}\right)\delta_{i3}$ results from the Boussinesq approximation where $g$ is the gravitational acceleration and $\delta_{ij}$ is the Kronecker delta operator. 
The flow is driven by a geostrophic forcing imposed through the net (pressure gradient $-$ Coriolis) body force $\tilde{F}_i = \left(\tilde{u}_2 - V_G \right)f_C \delta_{i1} - \left(\tilde{u}_1 - U_G \right)f_C \delta_{i2}$, where ($U_g$, $V_g$) is the geostrophic wind vector representing the synoptic pressure forcing and $f_C$ is the Coriolis frequency.

The deviatoric part of the SGS kinematic stress tensor ($\tau_{ij}$) is parameterized using an eddy-viscosity model: $\tau_{ij} = -2 \nu_T \tilde{S}_{ij}$, where $\nu_T = \left(C_{S} \Delta\right)^{2} |\tilde{S}|$ is the eddy viscosity, $\tilde{S}_{ij}$ the resolved rate of strain tensor, and $C_S$  the Smagorinsky coefficient computed dynamically using the Lagrangian scale-dependent dynamic approach of \citet{BouZeid2005}. The SGS heat flux $q_j$ is similarly modeled using an eddy diffusion approach, $q_{j} = -\left(\nu_T/Pr_{sgs}\right)\partial \tilde{\theta}/\partial x_j$, where the eddy diffusivity ($\nu_T/Pr_{sgs}$) is computed from $\nu_{T}$ using a constant SGS Prandtl number, $Pr_{sgs} = $ 0.4.  

The LES algorithm employs pseudo-spectral discretization in the horizontal directions with full dealiasing, and second-order centered finite differences in the vertical direction on a staggered grid; see \citet{BouZeid2004} for full details. 
Time integration is performed with a second-order Adams–Bashforth scheme, and incompressibility is enforced by solving a three-dimensional Poisson equation for pressure at each time step \cite{Chorin1968}. The equations are solved with periodic boundary conditions in the horizontal directions. 
At the surface, momentum and heat fluxes are related to the surface-air differences in velocity or temperature through a local wall model based on the Monin--Obukhov similarity theory, with velocities and temperatures filtered at twice the horizontal grid spacing and sampled at a distance from the wall equal to half the vertical grid spacing (on a staggered grid) \cite{BouZeid2005}. 
The aerodynamic roughness length is set to $z_0=0.1$m in all simulations described below, and the thermal roughness length is $z_{0s} = 0.1 z_{0}$.
The upper boundary is stress-free and impermeable, with a Rayleigh damping layer applied near the top of the domain to absorb gravity waves when an inversion layer is added.

% The neutral and unstable simulations were initialized by spinning up the model for 8 hours on a neutral stability case, followed by an additional 3-6 hours for the prescribed experimental conditions (Section \ref{sec:methods}-\ref{ssec:exp}). 
% The neutral and unstable simulations were initialized with a geostrophic wind of 5 m s$^{-1}$, with a constant temperature profile of 280 K up to 1000 m and without a capping inversion. 
% The model was first run for 8 hours on a coarse numerical grid of 100$^3$, and the outputs were then interpolated to a 200$^3$ grid and integrated for another 13 hours, discarding the first hour for training. 
% The initialization of the stable cases is fully detailed in \citet{Huang2013}. 
% Briefly, simulations were initialized with a geostrophic wind of 8 m s$^{-1}$, with a constant initial temperature of 265 K up to 100 m capped by a 0.01 K/m inversion aloft. 
% The equations were integrated for 10 hours, first on a coarse numerical grid of $80^3$ for 6 hours, then interpolated to a grid 162 $\times$ 162 $\times$ 160 and integrated for another 4 hours. 
% The first hour, at high-resolution, was used as additional spin-up for the finest eddies to form, and statistics were then computed from horizontal and temporal averages over the last 3 hours.
% Details on the prescribed experimental conditions are outlined in Section \ref{sec:methods}-\ref{ssec:exp}.

\subsection{LES Setup}
\label{ssec:exp}

Linear operators were derived under neutral, stable, and unstable conditions by varying homogeneous bottom boundary conditions. 
For each stability regime, a dedicated LES dataset was generated and used to solve for the linear operator that maps mean profiles to their corresponding fluxes. 
In the stable boundary layer case, we draw on the simulations of \citet{Huang2013} conducted to represent cases from the Global Energy and Water Cycle Experiment (GEWEX) Atmospheric Boundary Layer Study (GABLS) project, which are briefly summarized here. 
This dataset consists of six large-eddy simulations of the stable boundary layer, forced by steady surface cooling rates of –0.25, –0.5, –1.0, –1.5, –2.0, and –2.5 K h$^{-1}$, spanning weakly to strongly stable regimes. 
The computational domain extended 800m$\times$800m horizontally and 400m vertically, discretized at approximately 5m$\times$5m$\times$2.5m, thereby capturing most of the small-scale turbulent eddies that characterize stable flows. 
A constant geostrophic wind of 8 m s$^{-1}$ at 73$^\circ$N provided the background shear necessary to maintain turbulence under the different cooling conditions. 
Simulations were initialized with a geostrophic wind of 8 m s$^{-1}$, with a constant initial temperature of 265 K up to 100 m capped by a 0.01 K/m inversion aloft. 
The equations were integrated for 10 hours, first on a coarse numerical grid of $80^3$ for 6 hours, then interpolated to a grid 162 $\times$ 162 $\times$ 160 and integrated for another 4 hours. 
The first hour, at high-resolution, was used as additional spin-up for the finest eddies to form, and statistics were then computed from horizontal and temporal averages over the last 3 hours.
% In \citet{Huang2013} , the rates of cascade of energy from the resolved to the subgrid scales were validated against field experimental data from \cite{BOUZEID2010} to ensure that the resolved scale were captured accurately at the various stabilities.

In addition, a dataset for neutral stability was generated to verify the operator in the absence of heat flux and to provide a transitional case between stability regimes. 
This dataset employed a computational domain of 6000m$\times$6000m in the horizontal and 1000 m in the vertical, discretized at 30m$\times$30m$\times$5m resolution. 
The simulations were driven by a geostrophic wind of 5 m s$^{-1}$ at a latitude of 85$^\circ$N, and the neutrally stable case was integrated for approximately 5 hours following spin-up. 
For the unstable convective regime, a similar configuration was used, but with imposed surface heat fluxes of 50, 100, 150, 200, and 250 $W,m^{-2}$. 
A summary of all simulation parameters is provided in Table \ref{tab:exps}.

The neutral and unstable simulations were initialized with a geostrophic wind of 5 m s$^{-1}$, with a constant temperature profile of 280 K up to 1000 m and with a stress-free impermeable top boundary (representing a exceedingly strong capping inversion). 
The model was first run for 8 hours on a coarse numerical grid of 100$^3$, and the outputs were then interpolated to a 200$^3$ grid and integrated for another 13 hours, discarding the first hour as additional spin-up. 

\begin{table}
\resizebox{\columnwidth}{!}{%
\begin{tabular}{|l|c|c|c|c|c|c|}
\hline
                  & \textbf{$M_g$ (m/s)}& \textbf{Nx, Ny, Nz}& \textbf{Lx, Ly, Lz (m)} & \textbf{Boundary Condition}                              & \textbf{dt (s)} & \textbf{$f_c$ (rad/s)}\\ \hline
\textbf{Stable}   & 8                 & 162$\times$162$\times$160           & 800$\times$800$\times$400& $\partial \theta/\partial t |_{z=o} = -$ {[}0.5, 1, 1.5, 2, 2.5{]}$K h^{-1}$& 0.024       & 1.39E-4     \\ \hline
\textbf{Neutral}  & 5                 & 200$\times$200$\times$200           & 6000$\times$6000$\times$1000          & $\overline{w'\theta'}|_{z=0}$ = 0                                                & 0.05        & 1.45E-4     \\ \hline
\textbf{Unstable} & 5                 & 200$\times$200$\times$200           & 6000$\times$6000$\times$1000          & $\overline{w'\theta'}|_{z=0}=$  {[}50,100, 150, 200, 250{]} $Wm^{-2}$& 0.05        & 1.45E-04    \\ \hline
\end{tabular}%
}
\caption{LES experiment configurations for stable, neutral, and unstable ABL regimes.}
\label{tab:exps}
\end{table}

\section{Results}
\label{sec:res}

In this section, we present the derived parameterization models and compare their performance, \textit{a priori}, against the standard KPP scheme across all stability regimes at steady states. 
The parameterizations are then implemented in a single-column model (SCM) or SCM, to assess their reliability, \textit{ a posteriori} in an operational setting. 
\textit{A posteriori} tests involving abrupt transitions between stability regimes are also performed, followed by generalization tests in which interpolated operators are used to reconstruct flux profiles under unseen conditions. 
The two testing approaches are complementary and essential to verify that the model structure and assumptions are valid irrespective of their impact on the evolution of the ABL state (\textit{a priori}), and that the model is numerically stable (not guaranteed when the input mean states start evolving and affecting the model output) and yields accurate results in actual implementations in the RANS prognostic equations (\textit{a posteriori}). 
Table \ref{tab:experiments} provides an overview of the various tests and results.

\begin{table}[!htbp]
\centering
\caption{Summary of experiments discussed in Section~3.}
\label{tab:experiments}
\renewcommand{\arraystretch}{1.25}
{\small
\begin{tabularx}{\linewidth}{
>{\raggedright\arraybackslash}p{0.06\linewidth}
>{\raggedright\arraybackslash}p{0.10\linewidth}
>{\raggedright\arraybackslash}p{0.30\linewidth}
>{\raggedright\arraybackslash}p{0.38\linewidth}
>{\centering\arraybackslash}p{0.06\linewidth}
}
\toprule
Sec. & Test & Case / forcing & What is evaluated & Fig(s).\\
\midrule

3a & A priori
& Stable and unstable LES cases
& Learned univariate linear operators $(A_\theta, A_u, A_v)$ [Eq.~(8)] and their vertical (non)locality and scaling behavior as encoded by the operator structure.
& 1 \\

3a & A priori
& Stable example $\left.\partial\theta/\partial t\right|_{z=0}=-1~\mathrm{K\,h^{-1}}$; convective example $w'\theta'|_{z=0}=50~\mathrm{W\,m^{-2}}$
& Multivariate operators using $(\theta,u,v)$ jointly for each flux [Eq.~(9)]; flux reconstruction skill and localized reconstruction errors relative to LES.
& 2--3 \\

3b & A posteriori
& Stable ABL case (surface cooling; see Table~1)
& SCM time--height evolution of $(\theta,u,v)$ using multivariate learned operators [Eq.~(9)]; benchmark comparison against KPP [Eqs.~(11)--(14)].
& 4--5 \\

3b & A posteriori
& Convective ABL with $w'\theta'|_{z=0}=50~\mathrm{W\,m^{-2}}$
& SCM time--height evolution of $(\theta,u,v)$ using multivariate learned operators [Eq.~(9)]; agreement with LES and LES--SCM differences under convective forcing.
& 6 \\

3c & A posteriori
& Stable interpolation target $\left.\partial\theta/\partial t\right|_{z=0}=-1.5~\mathrm{K\,h^{-1}}$; interpolated from $-1.0$ and $-2.0~\mathrm{K\,h^{-1}}$
& Generalization via \emph{linear interpolation} between operators [Eq.~(8)] trained at neighboring forcings; comparison of LES, trained-operator SCM, and interpolated-operator SCM snapshots.
& 7 \\

3c & A posteriori
& Unstable interpolation target $w'\theta'|_{z=0}=100~\mathrm{W\,m^{-2}}$; interpolated from $50$ and $150~\mathrm{W\,m^{-2}}$
& Same interpolation test for convective conditions; comparison of instantaneous profile snapshots across LES, trained, and interpolated operators.
& 8 \\

3d & A posteriori
& Neutral initialization $\rightarrow$ (i) stable: $\partial\theta/\partial t=-1~\mathrm{K\,h^{-1}}$; (ii) unstable: $w'\theta'=150~\mathrm{W\,m^{-2}}$
& Robustness to step changes from neutral: smooth adjustment without spurious oscillations and physically consistent evolution of $(\theta,u,v)$, using mutivariate learned operators.
& 9 \\

3e & A posteriori
& Abrupt stability flips: unstable$\rightarrow$stable and stable$\rightarrow$unstable
& ``Stability-flip'' stress test: numerical stability and qualitative adjustment, including operator rescaling to different SCM grids, using mutivariate learned operators.
& 10 \\

\bottomrule
\end{tabularx}
}
\end{table}

\subsection{Linear Operators}
\label{ssec:linOps}

\begin{figure}
\noindent\includegraphics[width=\textwidth]{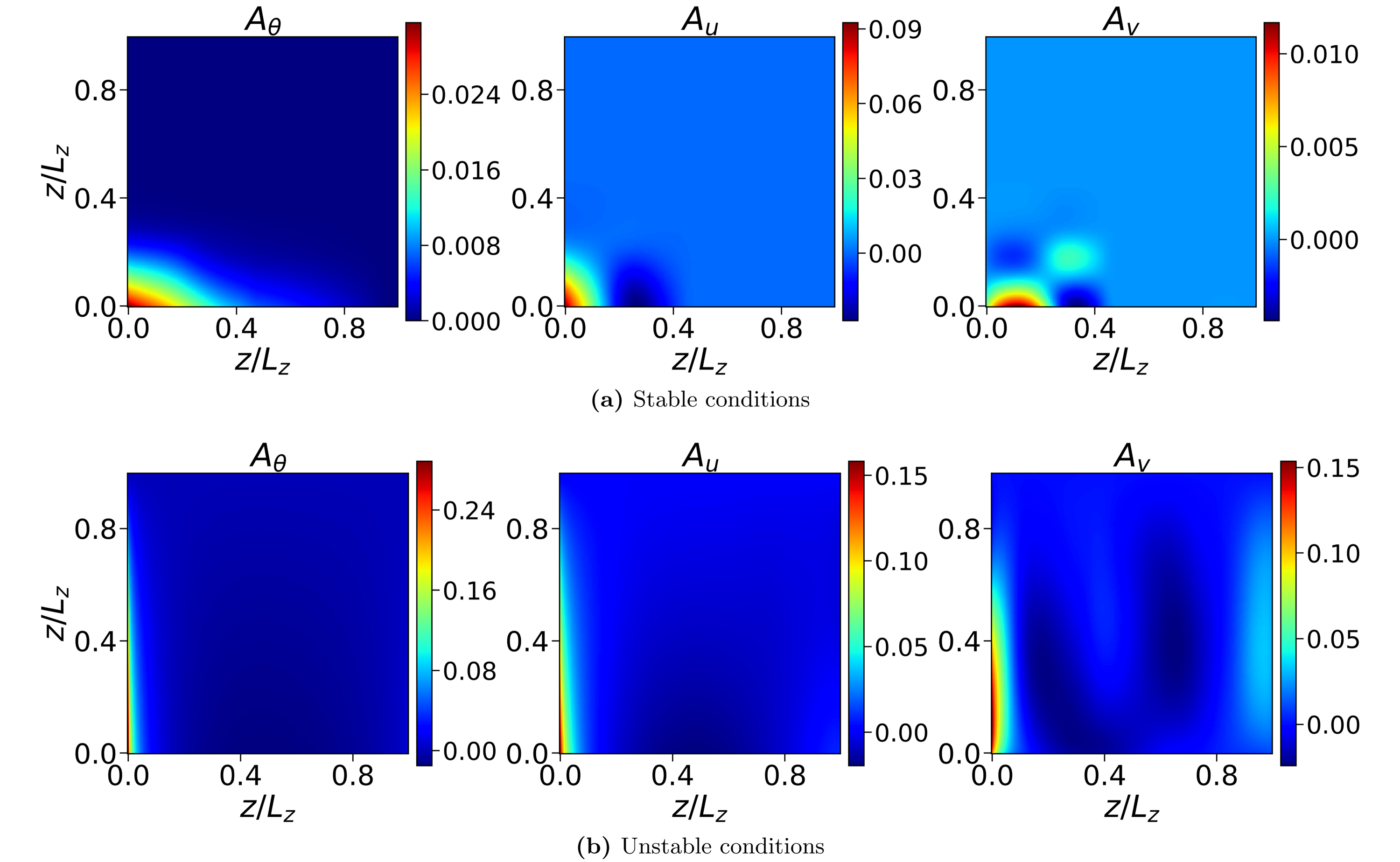}
\caption{Linear operators mapping mean profiles to fluxes under (a–c) stable and (d–f) unstable conditions. Each matrix row encodes an algebraic stencil resembling the overall operations, such as the vertical derivative. 
}
\label{fig:ops_simple}
\end{figure}

We first examine linear operator formulations of increasing complexity. 
All parameterizations were constructed using normalized input–output pairs and rely on first-order statistics only, i.e., mean profiles.  
The simplest class of parameterizations estimates each flux using only the transported scalar, expressed in a Galilean invariant form as:
% \pagebreak
\begin{equation}\label{eq:paramSep}
\begin{aligned}
\frac{\overline{w'\theta'}}{\overline{w'\theta'}(z=0)} &\approx A_{\theta} \frac{\overline{\theta} - \overline{\theta}(z=0)}{\overline{\theta}(z=L_z) - \overline{\theta}(z=0)}, \\
\frac{\overline{w'u'}}{\overline{w'u'}(z=0)} &\approx A_{u} \frac{\overline{u} - \overline{u}(z=0)}{\overline{u}(z=L_z) - \overline{u}(z=0)}, \\
\frac{\overline{w'v'}}{\overline{w'v'}(z=0)} &\approx A_{v} \frac{\overline{v} - \overline{v}(z=0)}{\overline{v}(z=L_z) - \overline{v}(z=0)}.
\end{aligned}
\end{equation}
Figure~\ref{fig:ops_simple} shows the corresponding operators mapping mean profiles of temperature, $u$-velocity, and $v$-velocity to their respective fluxes. 
Each operator matrix can be interpreted as a set of algebraic equations (rows) that effectively encode operations (e.g.,  spatial derivatives and multiplication by some eddy diffusivity for the local part of the fluxes), enabling the inputs to be mapped to their flux counterparts. 
Each row is multiplied by the mean vertical profile to compute the flux at that height, meaning each column encodes the influence of a given height on the rest of the ABL.

\begin{figure}
\noindent\includegraphics[width=\textwidth]{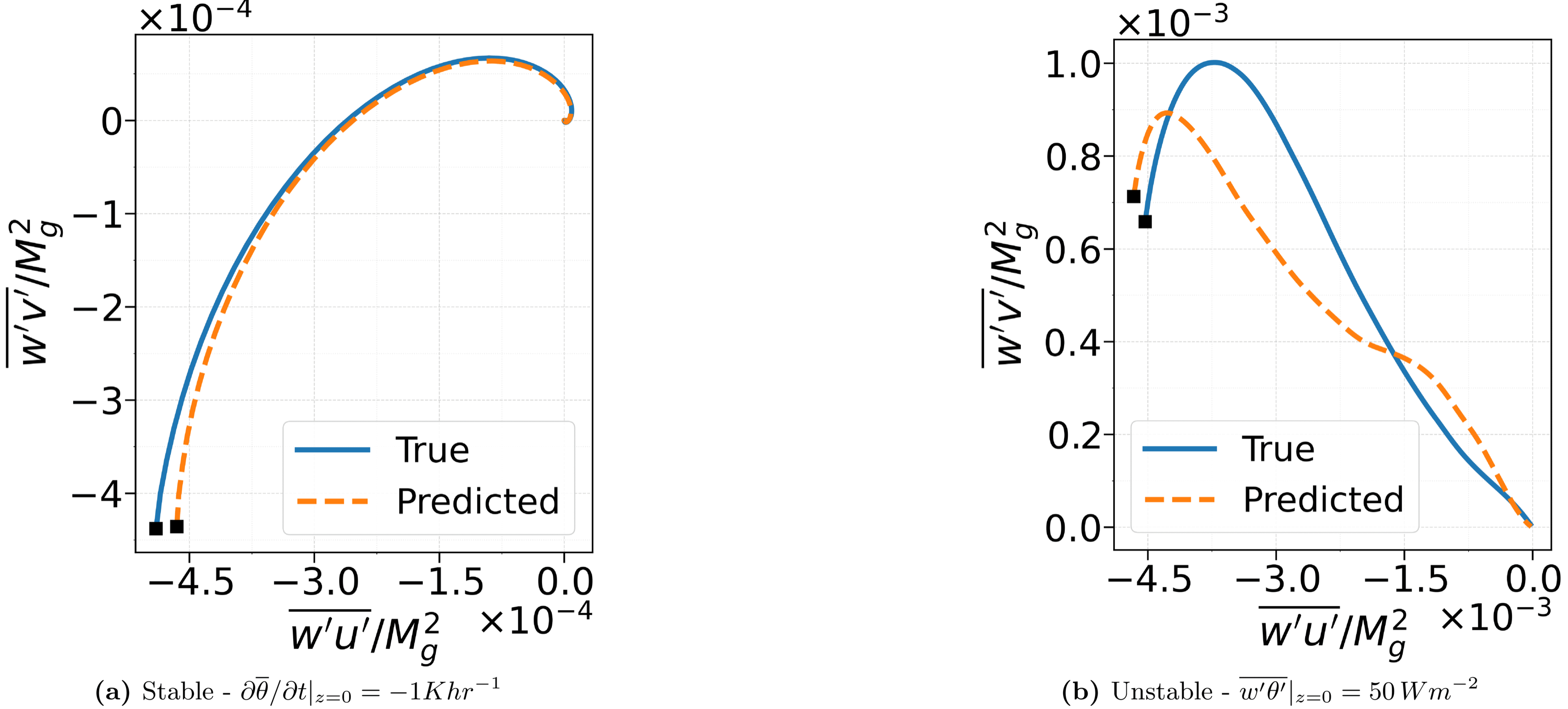}
\caption{Momentum flux hodographs showing the relation between the turbulent flux components $\overline{w'u'}$ and $\overline{w'v'}$ and its variability with height for (a) stable and (b) unstable boundary layer conditions. Solid blue curves denote LES reference values, while dashed orange curves correspond to predictions from the learned operators. Black markers indicate surface values.}
\label{figs:hodographs}
\end{figure}

For the stable boundary layer illustrated in Figures~\ref{fig:ops_simple}(a)–(c), the heat flux operator exhibits a sharp cutoff above the inversion height, visible in rows of zero coefficients, and a decaying magnitude of entries with increasing distance from the surface. 
This indicates mostly localized interactions inside the stable ABL with diminishing correlation away from the ground, but the fluxes across the stable ABL are still coupled to the surface (no $z$-less behavior). 
In contrast, the momentum operators display a tighter spatial locality, with non-zero contributions largely confined to within the boundary layer and a harmonic structure that encodes the Ekman turning, which is further portrayed in the momentum flux hodographs (Figure \ref{figs:hodographs}, left panel).  
In the unstable case, Figures~\ref{fig:ops_simple}(d)–(f), the operators span the entire boundary layer depth, with both the heat flux and $u$-momentum operators showing strong non-locality. 
The vertically-elongated structure indicates a strong influence of the surface across the full depth, representing the flux-carrying thermals rising from the surface.
Meanwhile, the $v$-momentum operator exhibits weaker oscillatory patterns, echoing the harmonic behavior of the stable case.

Figure \ref{figs:hodographs} illustrates the momentum–flux hodographs for the stable and unstable boundary layers to further illustrate the dynamical structure encoded by the learned operators.
In the stable boundary layer (Figure~\ref{figs:hodographs}a), the trajectory of $(\overline{w'u'},\overline{w'v'})$ follows a smooth, curved path, characteristic of Ekman-type turning, reflecting the progressive rotation of the turbulent stress vector with height. 
The predicted hodograph closely tracks the true LES trajectory, capturing both the curvature and the magnitude of the flux vector, with only minor deviations near the strongest momentum flux values close to the surface. 
In the unstable case (Figure~\ref{figs:hodographs}b), the hodograph exhibits a broader excursion and a more monotonic evolution, consistent with convection-driven weakened veer, which is typical of unstable conditions. 
The predicted trajectory again reproduces the overall shape of the LES curve, although the peak momentum flux magnitude is slightly underestimated and the veer is not captured as well as under stable conditions. Velocity hodographs, not shown here, also show very good agreement under stable conditions and larger errors under unstable ones.

To improve on these results, we increase the model complexity, such that all mean profiles ($\overline{\theta}$, $\overline{u}$, $\overline{v}$) are used jointly to predict each flux:
\begin{equation}\label{eq:block}
\begin{aligned}
\frac{\overline{w'\theta'}}{\overline{w'\theta'}(0)} &\approx A_{\theta \leftarrow \theta} \frac{\overline{\theta} - \overline{\theta}(0)}{\overline{\theta}(L_z) - \overline{\theta}(0)} + A_{\theta \leftarrow u} \frac{\overline{u} - \overline{u}(0)}{\overline{u}(L_z) - \overline{u}(0)} + A_{\theta \leftarrow v} \frac{\overline{v} - \overline{v}(0)}{\overline{v}(L_z) - \overline{v}(0)}, \\
\frac{\overline{w'u'}}{\overline{w'u'}(0)} &\approx A_{u \leftarrow \theta} \frac{\overline{\theta} - \overline{\theta}(0)}{\overline{\theta}(L_z) - \overline{\theta}(0)} + A_{u \leftarrow u} \frac{\overline{u} - \overline{u}(0)}{\overline{u}(L_z) - \overline{u}(0)} + A_{u \leftarrow v} \frac{\overline{v} - \overline{v}(0)}{\overline{v}(L_z) - \overline{v}(0)}, \\
\frac{\overline{w'v'}}{\overline{w'v'}(0)} &\approx A_{v \leftarrow \theta} \frac{\overline{\theta} - \overline{\theta}(0)}{\overline{\theta}(L_z) - \overline{\theta}(0)} + A_{v \leftarrow u} \frac{\overline{u} - \overline{u}(0)}{\overline{u}(L_z) - \overline{u}(0)} + A_{v \leftarrow v} \frac{\overline{v} - \overline{v}(0)}{\overline{v}(L_z) - \overline{v}(0)}.
\end{aligned}
\end{equation}

\begin{figure}
\noindent\includegraphics[width=\textwidth]{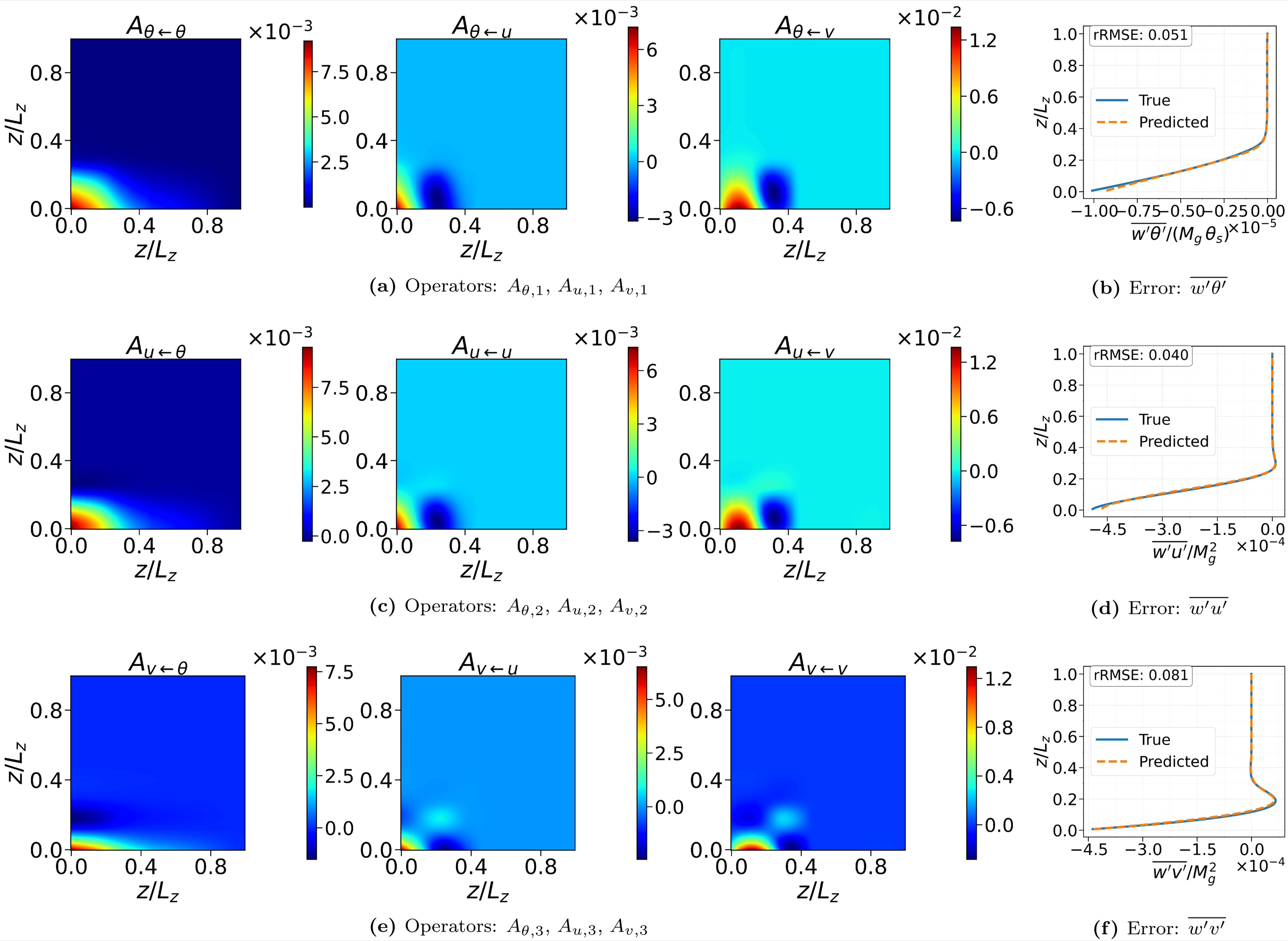}
\caption{multivariate operators for the stable ABL and their reconstruction skill for the stable boundary layer experiment ($\partial\theta/\partial t_{z=0}\!=\!-1\,Kh^{-1}$). Using $(\overline{\theta},\overline{u},\overline{v})$ jointly to predict each flux markedly reduces errors relative to scalar-only operators and better captures nonlocal interactions.}
\label{fig:ops_complex_stable}
\end{figure}

\begin{figure}
\noindent\includegraphics[width=\textwidth]{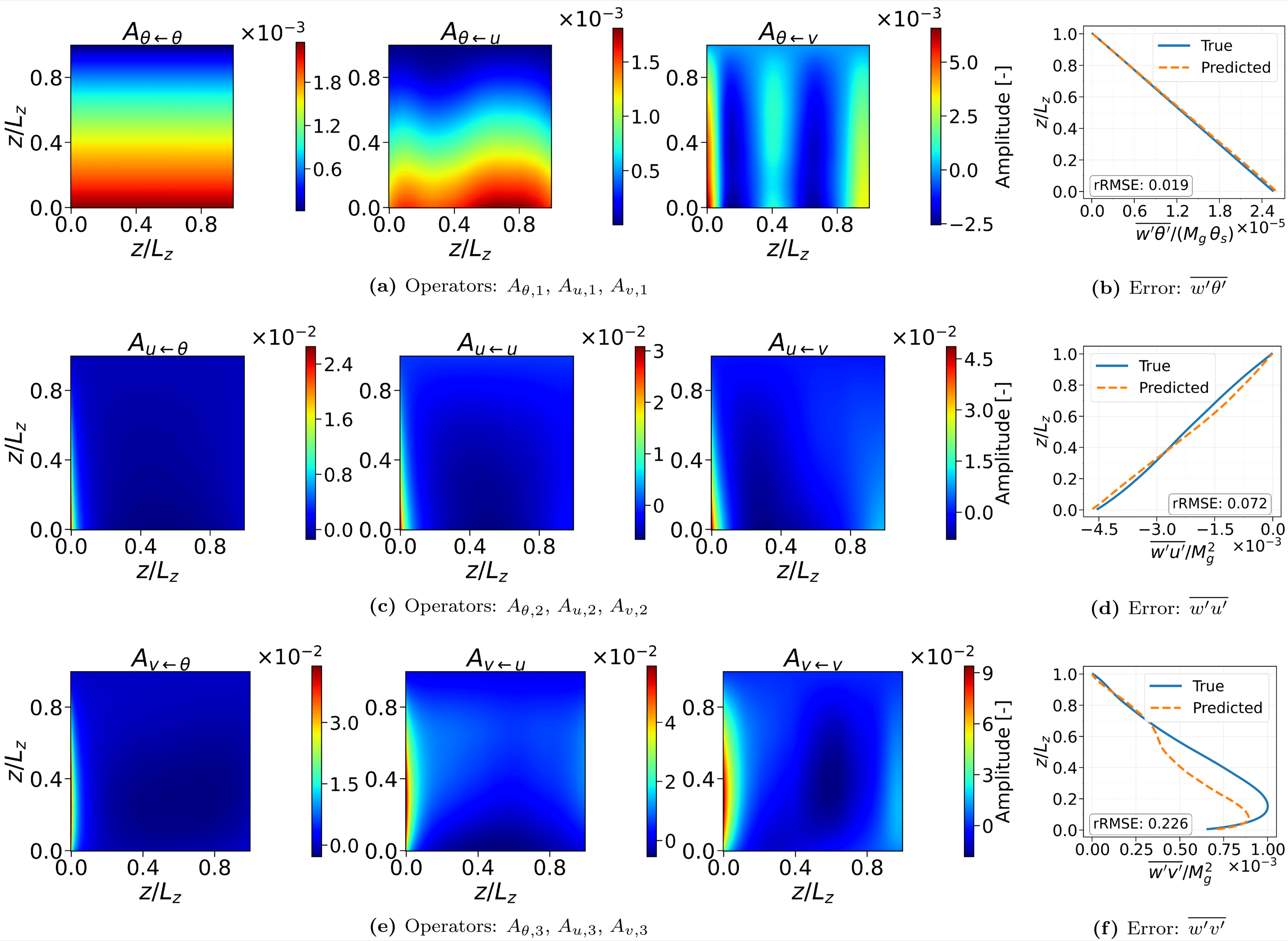}
\caption{Multivariate operators for the convective ABL and their reconstruction skill for the convective boundary layer experiment ($\overline{w'\theta'}_{z=0}=50\,W/m^2$). 
  Using $(\overline{\theta},\overline{u},\overline{v})$ jointly to predict each flux reduces errors relative to scalar-only operators and better captures nonlocal interactions.}
\label{fig:ops_complex_unstable}
\end{figure}

Figures~\ref{fig:ops_complex_stable} and \ref{fig:ops_complex_unstable} show representative operators and their ability to reconstruct LES-derived fluxes for the stable and unstable cases, respectively.
Compared to the scalar-only formulation, shown in the Supplementary, the inclusion of all mean fields as inputs reduces reconstruction errors and captures nonlocal interactions more effectively.
The error reduction is particularly significant for the $u$ flux profile; under stable conditions root mean square error decreased from 0.233 to 0.04, and comparable results under unstable conditions from 0.048 to 0.072.
Errors are localized to dynamically complex regions—near the surface and inversion height in stable regimes, near the surface and entrainment zone in unstable regimes, and to flux components with small magnitude ($\overline{w^\prime v^\prime}$ under unstable conditions).
This underscores the advantage of multivariate operators, particularly in regimes with strong coupling between heat and momentum fluxes, where conventional closure would use a Richardson number modification to the turbulent viscosities and diffusivities.

\subsection{Single Column Model Set-up}
\label{ssec:SCM}

The flux parameterizations derived in the previous section were next tested in an \textit{a posteriori} SCM framework. 
In this setting, the mean profiles evolve under the governing equations:
\begin{equation}\label{eq:scm}
    \begin{aligned}
        \frac{\partial \overline{\theta}}{\partial t} =& - \frac{\partial}{\partial z} (\overline{w'\theta'}), \\
        \frac{\partial \overline{u}}{\partial t} =& - \frac{\partial}{\partial z} (\overline{w'u'}) + f_c (\overline{v} - V_g), \\
        \frac{\partial \overline{v}}{\partial t} =& - \frac{\partial}{\partial z} (\overline{w'v'}) - f_c (\overline{u} - U_g),
    \end{aligned}    
\end{equation}
all variable are defined as in priori sections.
This \textit{a posteriori} setting enables evaluation of the parameterizations in the presence of online feedback from the fluxes to the evolving state. 
The simulations here are integrated from the initial conditions of the LES corresponding to the case studied. 

\begin{figure}
\noindent\includegraphics[width=\textwidth]{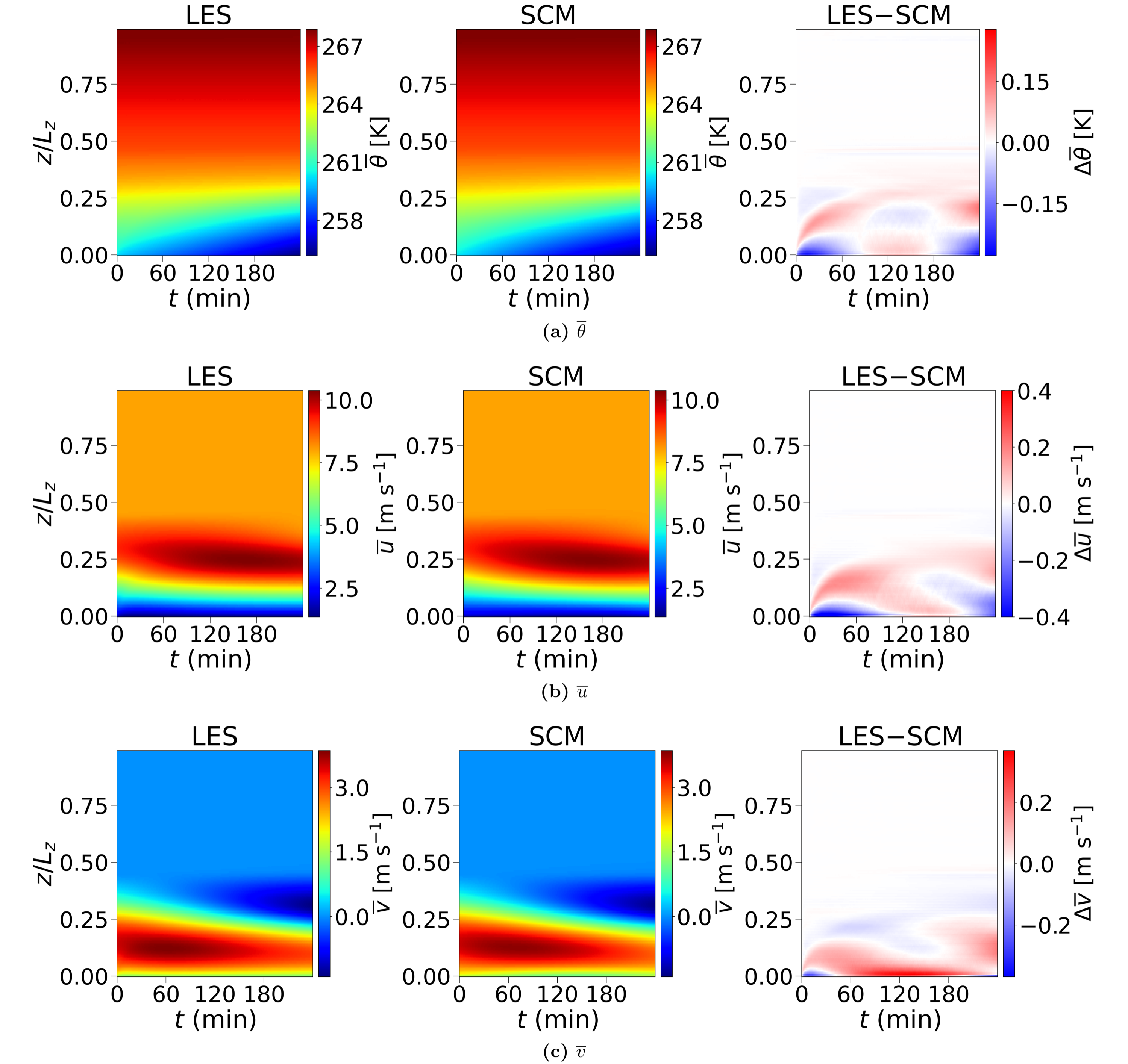}
\caption{Time-height plots of (a) temperature, (b) $u$-velocity, and (c) $v$-velocity for the stable boundary layer case, comparing LES (left), \textit{a posteriori} SCM using the data-driven parameterization (center), and their difference (right). 
The parameterization tracks the LES to within $\pm 1$ K for temperature and $\sim 10\%$ of geostrophic wind for velocities, while reproducing canonical features such as Ekman turning. Results are shown for the multivariate parameterization of Equation \ref{eq:block}.}
\label{fig:scm_stable}
\end{figure}

Figure \ref{fig:scm_stable} presents time-height (Hovm\"oller) plots for the temperature, u-velocity, and v-velocity components predictions of the LES and the \textit{a posteriori} SCM, and their difference are shown for the parameterization of Equation~\ref{eq:block}. 
The plots demonstrate that the \textit{a posteriori} solution remains close to the LES: temperature errors remain within $\pm 1$ K and velocity errors within $\sim 10\%$ of the geostrophic wind. 
The parameterization also captures Ekman turning, confirming that the learned operators provide physically consistent state evolution under feedback.  

\begin{figure}
\noindent\includegraphics[width=\textwidth]{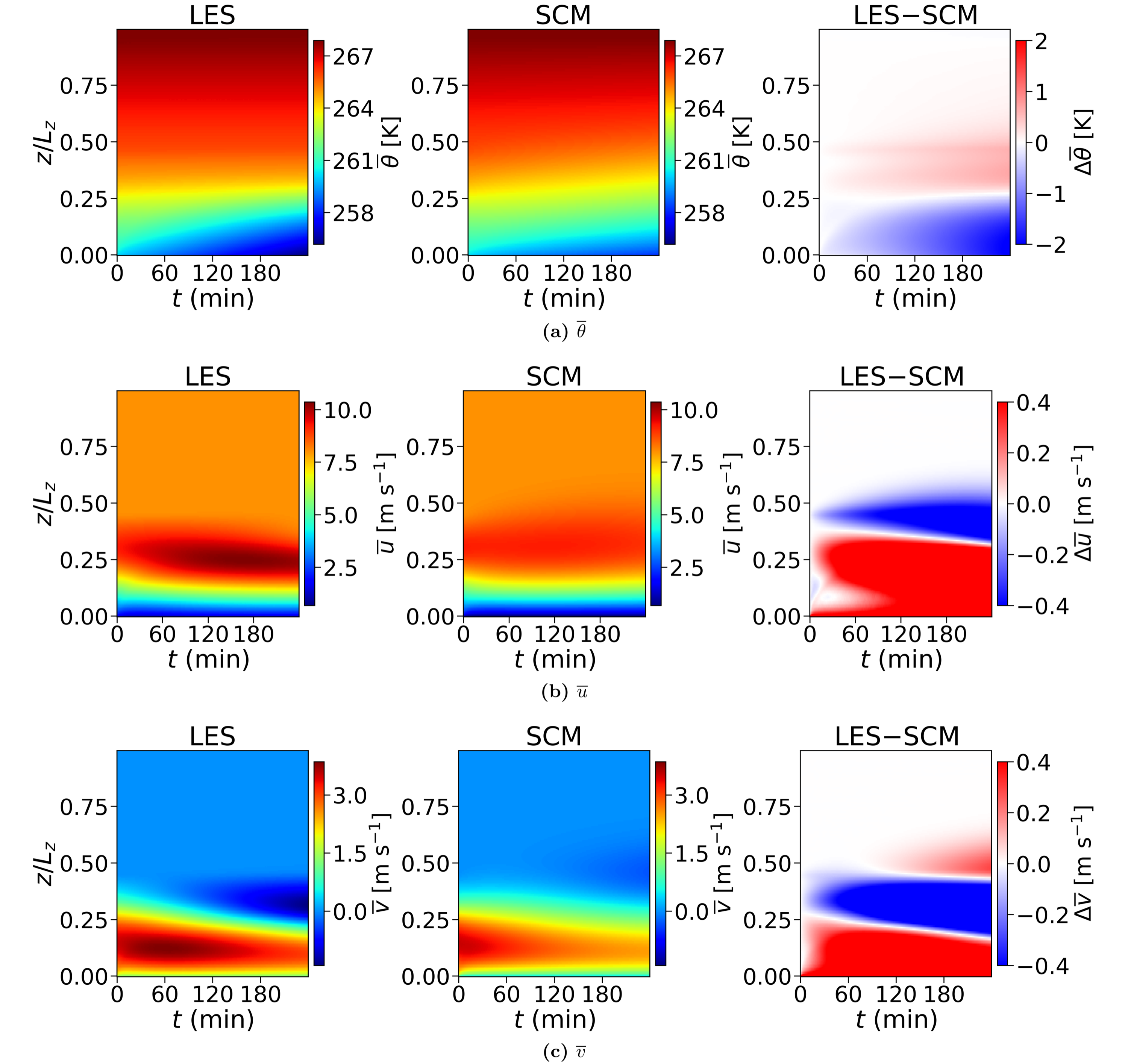}
\caption{Time-height plots of (a) temperature, (b) $u$-velocity, and (c) $v$-velocity for the stable boundary layer using the KPP scheme.}
\label{fig:kpp_stable}
\end{figure}

To benchmark against classical schemes, Figure~\ref{fig:kpp_stable} shows equivalent time-height plots for a KPP model based on the Troen–Mahrt parameterization \cite{TroenMahrt1986} with a Blackadar mixing length \cite{Blackadar1962}. 
The KPP model is thoroughly described in Appendix \ref{app:kpp}. 
The plots illustrates that the KPP profiles have much larger errors that exceed 2 K for temperature, and are more than twice larger than the linearized convolution errors for velocity.

\begin{figure}
\noindent\includegraphics[width=\textwidth]{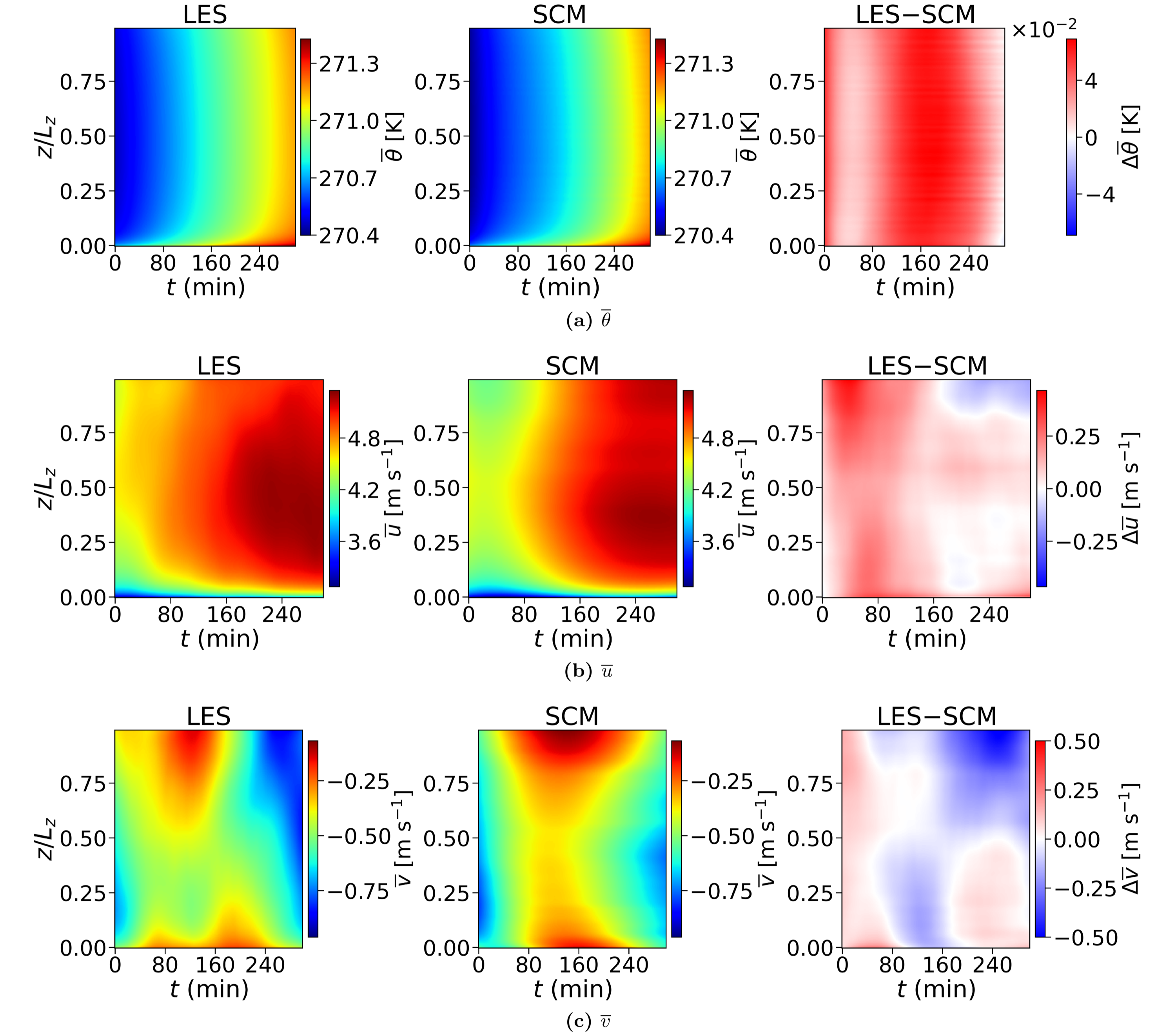}
\caption{Time-height plots of (a) potential temperature, (b) $u$-velocity, and (c) $v$-velocity for the unstable boundary layer ($\overline{w'\theta'}|_{z=0}=50\,W\,m^{-2}$), comparing (left) LES, (middle) \textit{a posteriori} SCM results through (right) their difference.}
\label{fig:scm_unstable}
\end{figure}

For the unstable boundary layer with $\overline{w'\theta'}|_{z=0}=50\,W\,m^{-2}$, Figure~\ref{fig:scm_unstable} compares LES and operator-based \textit{a posteriori} results. 
The data-driven parameterization captures all key features of the convective ABL: temperature is reproduced within $\pm 0.5$ K, slightly underestimated, while $u$- and $v$-velocities match LES hotspots of acceleration and deceleration within $\pm 10\%$. 
These results suggest that linear operators suffice for unstable regimes at first order, though nonlinear extensions (e.g., neural networks) may further improve fidelity.  
For brevity, the KPP results for this case are presented in the Supplementary. 
In summary, KPP reproduces large-scale patterns during the initial prediction period but loses skill quickly, with errors far exceeding those of the operator-based approach.

% SCM or profiles for the cases when we interpolate the linear operator to run different cases of cooling rates or bottom heat fluxes
\subsection{Generalizability Across Boundary Conditions}
\label{ssec:interpolation}

The generalizability of the derived parameterizations was further assessed by testing whether operators could be linearly interpolated across different surface-forcing conditions. 
Specifically, univariate operators derived for distinct surface cooling or heating rates were linearly combined, and the interpolated operators were then applied in the \textit{a posteriori} SCM setting. 
Their performance was compared against both the LES and the directly learned operator for the given condition.  

\begin{figure}
\noindent\includegraphics[width=\textwidth]{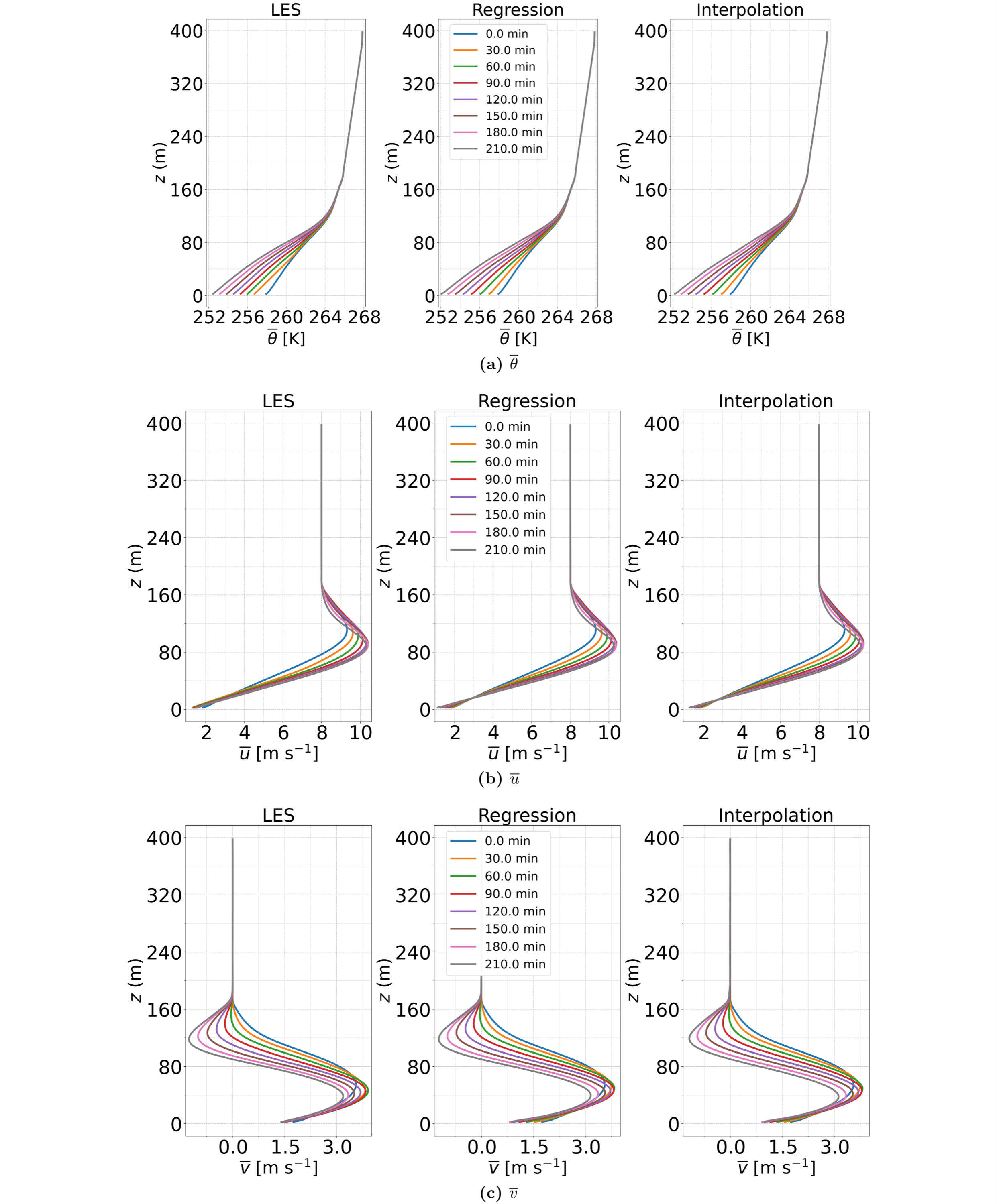}
\caption{Instantaneous snapshots of temperature, u-velocity and v-velocity profiles at select times for (a) LES data, (b) learned operator, and (c) interpolated operator for the stable ABL with $\partial \theta/\partial t|_{z=0} = -1.5\,K\,h^{-1}$.}
\label{fig:scm_stable_interp}
\end{figure}

Figure~\ref{fig:scm_stable_interp} presents results for the stable boundary layer, where the (left) LES case with $\partial \theta/\partial t|_{z=0} = -1.5\,K\,h^{-1}$ is compared against (middle) the operator trained on this condition, and (right) the interpolated operator (Equation~\ref{eq:paramSep}) obtained from the $-1.0$ and $-2.0\,K\,h^{-1}$ cases. 
Both operators closely reproduce the LES solution: the temperature inversion height and near-surface gradient are well matched, while the $u$- and $v$-velocity components capture the low-level jet. 
This demonstrates that interpolated operators retain the essential physics of the stable ABL and provide results nearly indistinguishable from directly trained operators.  
Practically, this is a critical feature of the model since it implies that only a few normalized operators are required, spanning stable and unstable conditions, and that these operators can then be scaled by the boundary conditions (surface heat flux in this case) to simulate any stability.

\begin{figure}
\noindent\includegraphics[width=\textwidth]{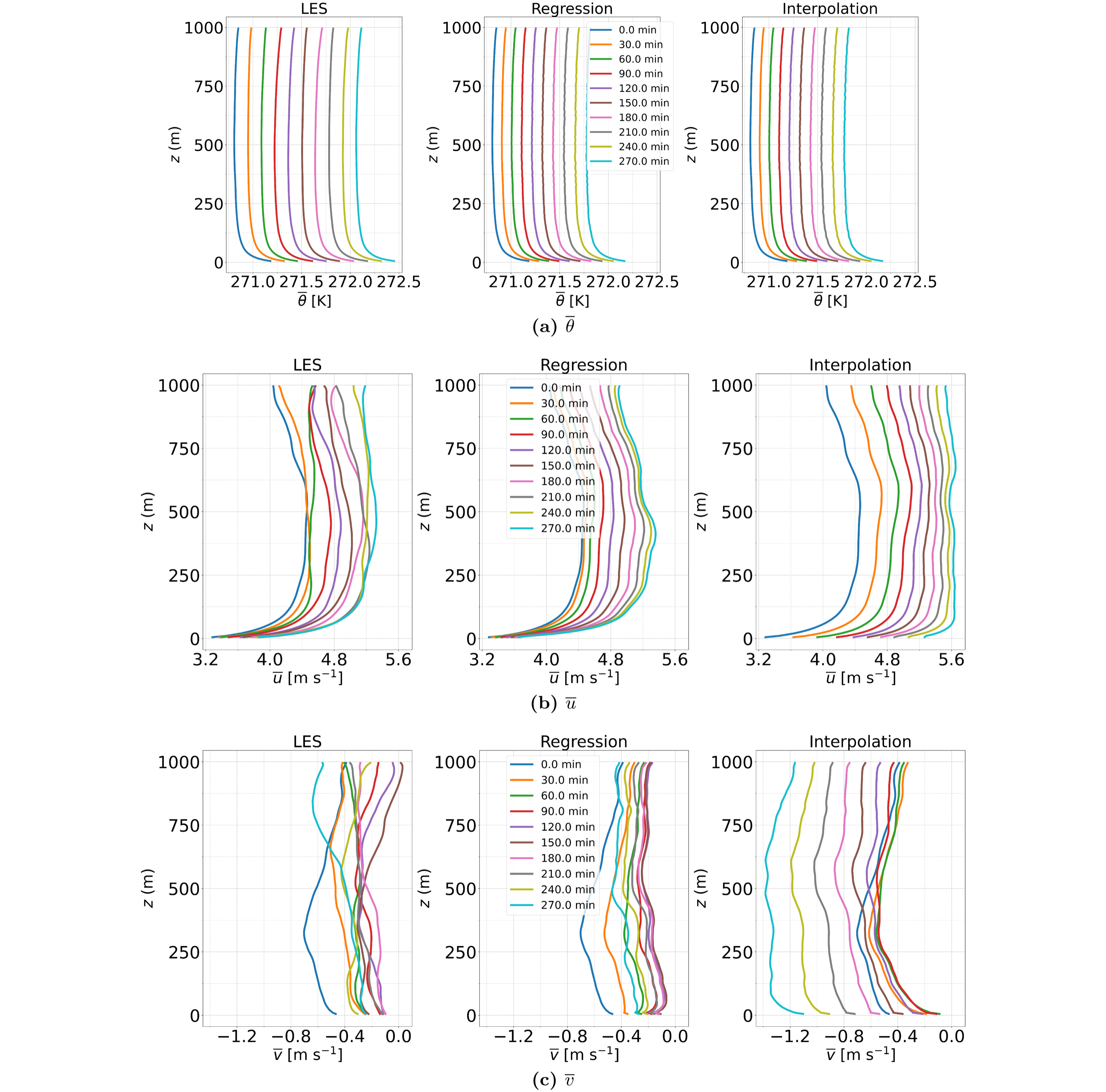}
\caption{Instantaneous snapshots of temperature, u-velocity and v-velocity profiles at select times for (a) LES data, (b) learned operator, and (c) interpolated operator for the unstable ABL with $\overline{w'\theta'}|_{z=0}=100\,W\,m^{-2}$.}
\label{fig:scm_unstable_interp}
\end{figure}

A similar test was conducted for the unstable boundary layer. 
Figure~\ref{fig:scm_unstable_interp} shows results for the (left) LES case with $\overline{w'\theta'}|_{z=0}=100\,W\,m^{-2}$, compared against (middle) the multivariate operator trained at this flux, and (right) an interpolated operator obtained from the $50$ and $150\,W\,m^{-2}$ cases. 
The interpolated profiles closely follow the LES and trained-operator solutions: temperature is reproduced within $\pm 1$ K, while velocity profiles capture the major features of convective acceleration and deceleration, with errors typically within $20\%$. 
Minor discrepancies appear near the surface and at the entrainment zone, where the nonlinear effects are strongest, but the interpolated operator still yields physically meaningful states.

% SCM outputs for the SCM transitions from:
% neutral -> unstable
% neutral -> stable
\subsection{Transition from Neutral Stability}
\label{ssec:scm_neutralIC}

In general circulation and operational models, atmospheric stability can undergo sudden transitions, which may induce numerical and physical instabilities in state estimates. 
To evaluate the robustness of the proposed parameterization, we perform a strenuous test by initializing the \textit{a posteriori} SCM from neutral stability and then transitioning to either stable or unstable surface boundary conditions, with fluxes estimated using the multivariate learned linear operators.  

\begin{figure}
\noindent\includegraphics[width=\textwidth]{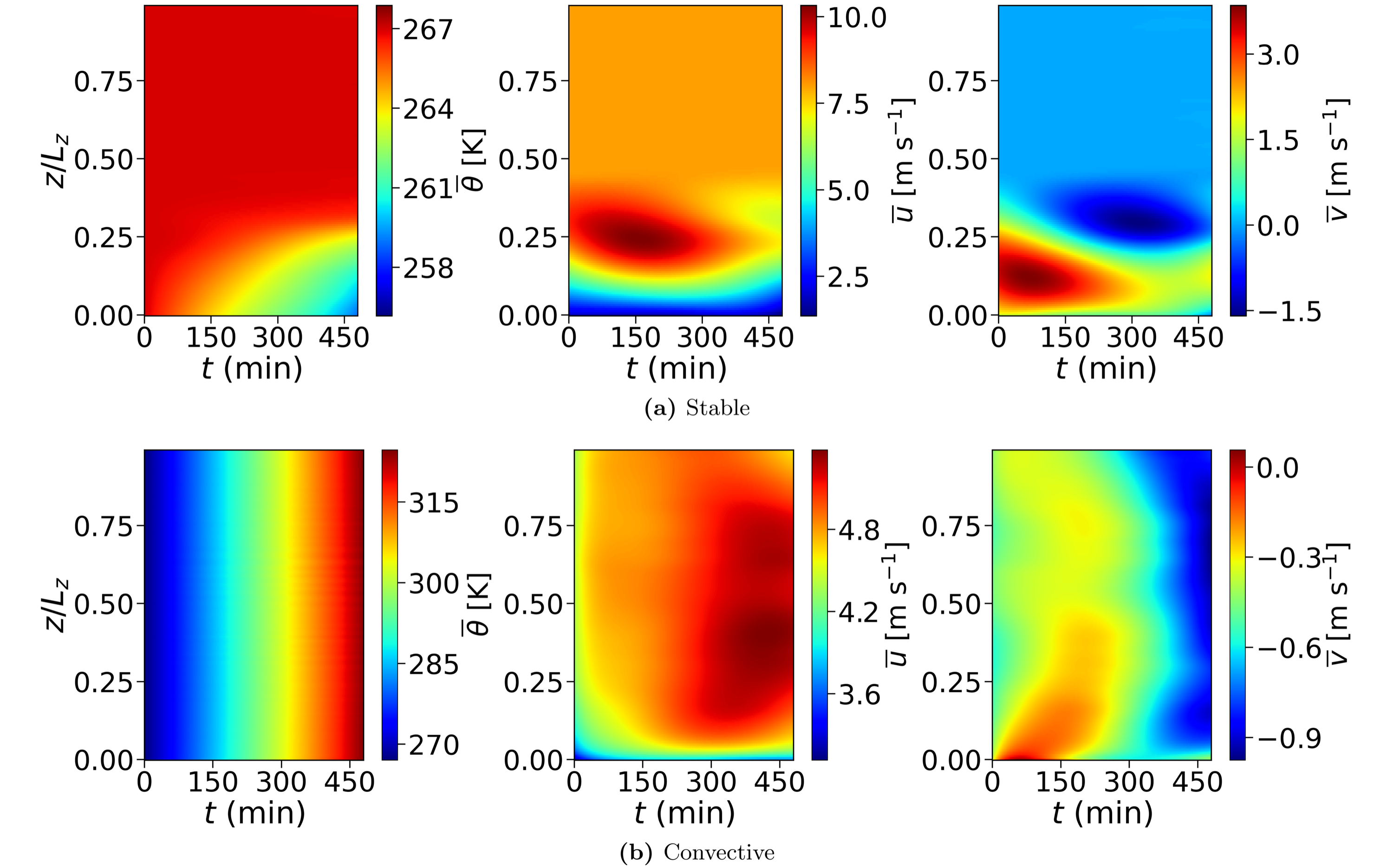}
\caption{Time-height plots of temperature, $u$-, and $v$-velocity for \textit{a posteriori} tests initialized from neutral stability. Shown are the cases for (a) transition to a stable ABL with $\partial \theta/\partial t = -1\,K\,h^{-1}$, and (b) transition to an unstable ABL with $\overline{w'\theta'}=150\,W\,m^{-2}$.}
\label{fig:scm_neutral_init}
\end{figure}

Figure~\ref{fig:scm_neutral_init} shows time-height plots for temperature, $u$-, and $v$-velocity components during these transitions. 
In the stable case, Fig.~\ref{fig:scm_neutral_init}(a), the temperature profiles evolve smoothly, with surface cooling producing a shallow inversion whose strength increases over time. 
The $u$- and $v$-velocities display inertial oscillations, manifesting as harmonic variations accompanied by low-level jet formation.  

In the unstable case, Fig.~\ref{fig:scm_neutral_init}(b), constant surface heating of $\overline{w'\theta'}=150\,W\,m^{-2}$ drives rapid development of a convective mixed layer, with temperatures increasing uniformly and enhanced warming near the surface. 
Although the long integration period (8 hours) with constant positive heat flux leads to unrealistically high absolute temperatures, the qualitative evolution remains physically consistent. 
Further studies on transient and time-dependent forcing (diurnal cycles) will be left for future work. 
The $u$- and $v$-velocities exhibit structures that closely resemble those of the unstable ABL in Section~\ref{ssec:SCM}, confirming that the parameterization transitions smoothly from neutral to convective conditions without introducing spurious oscillations or instabilities.

% SCM outputs for the SCM transitions from:
% stable   -> unstable
% unstable -> stable
\subsection{Transition Across Stability Regimes}
\label{ssec:scm_stabilityICs}

To probe robustness under abrupt regime changes, we perform ``stability-flip'' experiments in which the surface forcing is stepped to drive the ABL between unstable and stable states. 
We examine both directions, unstable$\rightarrow$stable and stable$\rightarrow$unstable, and diagnose the response of the \textit{a posteriori} SCM using the learned linear operators. 
The experiments were initialized from the LES profiles for stable ($\partial\theta/\partial t|_s=-1\,K\,hr^{-1}$) and unstable ($\overline{w'\theta'}|_s=100\,W\,m^{-2}$) conditions.
In all cases, the operators are pre-conditioned by a simple lifting/projection so they can be applied at a resolution different from that used for training.
In particular, the present study relies on simple interpolation operations to rescale the operators to the appropriate numerical grid.

\begin{figure}
\noindent\includegraphics[width=\textwidth]{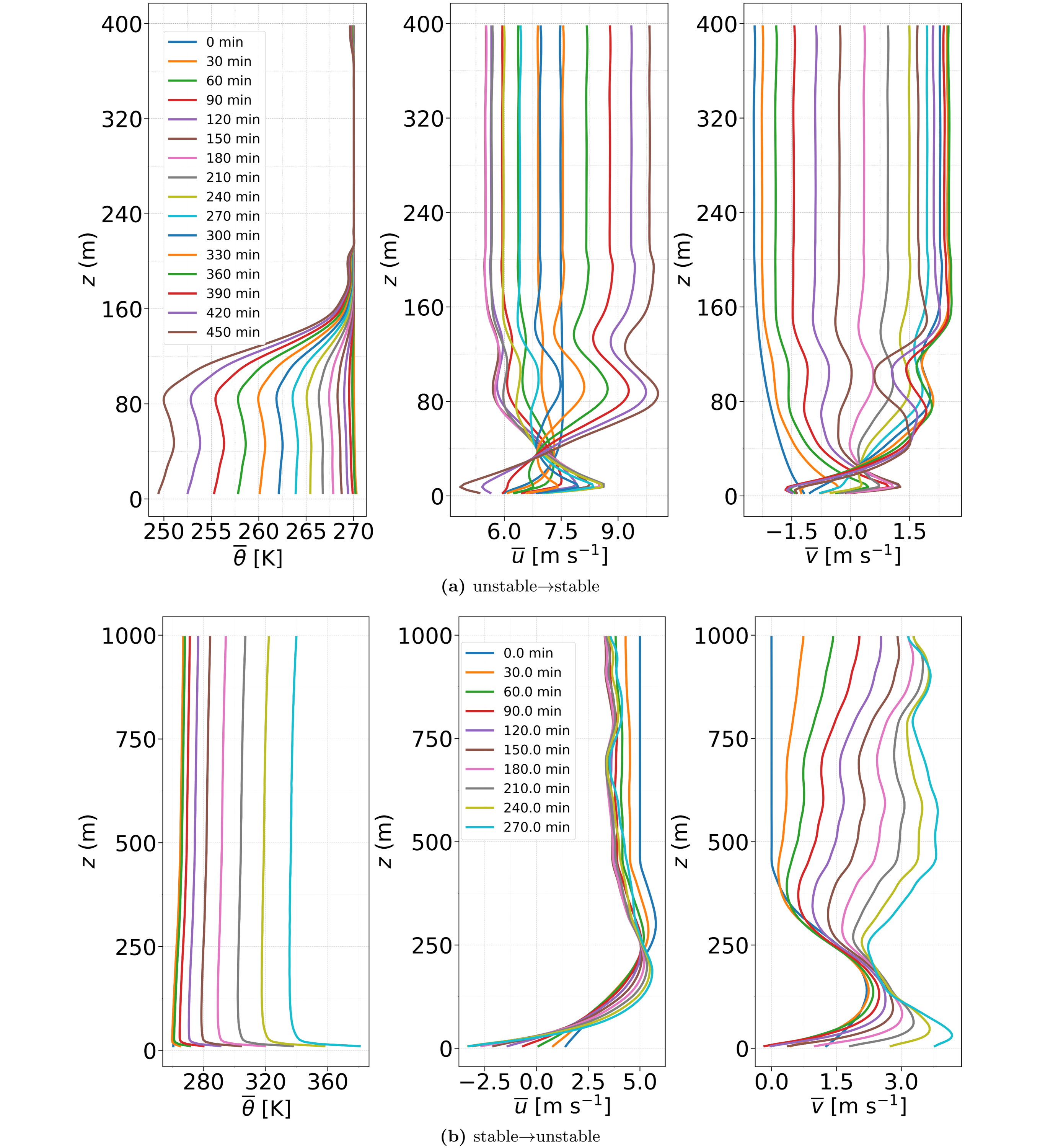}
\caption{Snapshots of potential temperature and mean $u$/$v$ during abrupt flips of surface forcing. (a) Unstable $\!\rightarrow\!$ stable: progressive surface cooling, growth of a capping inversion, and emergence of a low-level jet with inertial oscillations within $\sim$3 h. (b) Stable $\!\rightarrow\!$ unstable: rapid inversion breakup, formation and deepening of a convective mixed layer, and inertially modulated wind adjustment with a transient wind maximum near $0.2$–$0.3\,L_z$. In both directions, the \textit{a posteriori} SCM with learned operators remains numerically stable and reproduces the expected adjustment pathways.}
\label{fig:scm_stabSwitch}
\end{figure}

Figure~\ref{fig:scm_stabSwitch} shows snapshots of potential temperature and mean winds for (a) unstable$\rightarrow$stable and (b) stable$\rightarrow$unstable flips. 
The \textit{a posteriori} integrations remain numerically stable and free of spurious artifacts in both experiments.
Figure \ref{fig:scm_stabSwitch}(a) examines the case for transitions from unstable$\rightarrow$stable.
Following the step reduction in surface buoyancy flux, the state adjusts smoothly toward a nocturnal stable layer. 
Within $\sim\!3$~h, a low-level jet develops together with inertial oscillations in the horizontal winds, while the temperature profiles cool near the surface and develop a positive vertical gradient capped by an inversion. 
This sequence of cooling, shear intensification, low-level jet formation, and inertial oscillation is consistent with classical stable ABL adjustment.

Figure \ref{fig:scm_stabSwitch}(b) examines the case for transitions from stable$\rightarrow$unstable.
Imposing a positive surface heat flux (200$W/m^2$) rapidly breaks the inversion and establishes a convective mixed layer. 
Temperature profiles become nearly vertical below a sharpening capping inversion, and the time-height of $\overline{\theta}$ shows monotonic warming. 
Momentum adjusts more slowly: enhanced turbulent mixing entrains higher-momentum air downward, transiently reducing near-surface shear and producing a short-lived wind maximum around $0.2$–$0.3\,L_z$ that ascends as the layer deepens. 
The $\overline{u}$ and $\overline{v}$ fields display inclined bands and height-dependent veering—signatures of inertially modulated geostrophic adjustment. 

\section{Discussion}
\label{sec:discussion}

The results presented here suggest that the learned linear operators are effective closure models for coarse atmospheric simulations.
In operational terms, the proposed parameterization can be used in the same role traditionally occupied by K-profile or EDMF-type closures: given resolved mean profiles of $\overline{\theta}$, $\overline{u}$, and $\overline{v}$ on a vertical column, the operator returns the corresponding turbulent heat and momentum fluxes, whose vertical divergences then drive the prognostic equations of the mean state.
In this sense, the method is suited for \textit{a posteriori} application in single column models or fully three-dimensional numerical weather prediction and Earth systems models, although additional testing is still needed to assess the operation skill of the new closure.

A key advantage of the present framework is that it offers an interpretable middle ground between empirical physics-based closures and more opaque black-box machine-learning models.
The learned operator can be read directly as a map from the resolved mean state to the unresolved flux.
Its coefficient structure provides physical information that is usually hidden in purely data-driven surrogates. 
In particular, near-diagonal dominance corresponds to mostly local downgradient flux behavior, whereas broad off-diagonal support suggests nonlocal transport.
Classical parameterizations then arise as limiting cases of the learned model, where a purely local eddy diffusivity closure would correspond to an operator concentrated around the diagonal, whereas broader kernels recover the integral influence of vertical levels, which is a characteristic of nonlocal mass transport schemes.

The operators also enable an interpretable view of the parameterization. 
For instance, the stable boundary layer, the operators are comparatively localized, with heat-flux coefficients that weaken rapidly away from the surface and effectively vanish above the inversion.
This structure is consistent with turbulence that is confined to a shallow layer, which is almost decoupled from the free atmosphere, and strongly constrained by stratification.
Momentum operators are also localized and display oscillatory patterns associated with Ekman turning.
In contrast, the unstable operators exhibit support across most of the boundary-layer depth, which reflects the fact that convective plumes connect surface forcing to a much larger vertical extent.

Overall, the main contribution of this study is that a practical next generation of boundary layer closures may not require abandoning physical structure in favor of fully black-box models. 
Instead, the present results support a different direction, whereby one learns the closure as an operator constrained by the governing column dynamics, inspects its structure for physical meaning, and use that structure to guide both implementation and theory. 
The present framework is valuable because it offers a new language for interpreting turbulent transport and improves predictive skill relative to a standard KPP benchmark. 
The learned operators identify when the closure should behave locally, when it must behave nonlocally, and how thermal and momentum fields become dynamically coupled across stability regimes.

From an implementation standpoint, these operators could be distributed alongside legacy boundary layer codes in weather and climate models as precomputed coefficient tables. 
In practice, the online evaluation of the closure reduces to applying a small set of matrix--vector products to the resolved column profiles, followed by the same flux divergence update already performed in existing turbulence schemes. 
This makes the method attractive for operational deployment, since it can be inserted into established physics packages with minimal modification to the host solver and with limited additional computational overhead. 
One can broadly envision operational systems using a library of operators indexed by stability regime or surface forcing. 
Interpolation between nearby operators, and to the coarser resolutions of the parent model, would then provide a simple pathway for implementation across a range of atmospheric conditions.

\section{Conclusion}
\label{sec:conc}

We developed a data-driven framework for turbulent flux parameterization across convective to stable conditions in the atmospheric boundary layer using operator learning. 
Large-eddy simulations (LES) were used to learn global mappings from mean profiles ($\bar{\theta}$, $\bar{u}$, $\bar{v}$) to vertical fluxes (heat and momentum). 
Starting from a reverse-engineering of classical closures, we recast flux estimation as a convolution-type operation between an operator and a mean profile. 
Linearizing this relation reduces the task to learning a linear operator that acts on normalized mean fields, yielding a transparent and physics-interpretable parameterization without prescribing its functional form \textit{a priori}.
This methodology enables learning the ``best" parameterization, as learned from high-resolution LES data, without additional assumptions on the structure of the parameterization. 

\textit{A priori} and \textit{a posteriori} assessments show that the learned operators recover LES-consistent fluxes and states with high reliability. 
The proposed parameterization was demonstrated to accurately recover the solution from the validation dataset, indicating that the learned operators are reliable for estimating turbulent fluxes. 
The learned operators show similar features for the same stability condition, where the magnitude of the coefficients increases with increasing stability strength. 
Specifically, the learned operators indicate strong non-locality where virtually all numerical grid points are affected by the surface conditions, even under statically-stable regimes, with a decreasing effect with vertical distance. 
This provides new insight towards enhancing our conceptual understanding and physics-based parameterizations through inspiration drawn from data. 

In \textit{a posteriori} settings, the horizontally-averaged equations are solved, where fluxes are approximated using the derived linear operator, enforcing feedback to the reduced-order dynamical system. Despite this feedback, the data-driven parameterization remains numerically stable even with abrupt regime transitions, and faithfully reproduces the canonical features of the stable and unstable ABLs. 
In stable cases, it yields surface-based cooling with a retained capping inversion and an inertially-modulated low-level jet with realistic veering/backing, with temperature errors typically within $\pm 1$~K and velocity errors $\sim 10\%$ of geostrophic wind. 
In convective cases it captures rapid surface warming, mixed-layer deepening with entrainment warming, and momentum mixing toward vertically uniform winds, with temperature errors $\lesssim 0.5$–$1$~K and velocity errors $\sim 10$–$20\%$. 
Compared against a standard K-profile parameterization (KPP), the operator-based scheme shows systematically smaller, more localized errors and maintains skill over longer horizons.

To assess the generalizability of the proposed method, we derive several operators and apply simple linear interpolation across the operators learned for different surface conditions, and apply the interpolated operator in an \textit{a posteriori} test. 
Across both stability regimes, interpolating the learned operators between different surface conditions yields heat and momentum flux estimates that drive LES-consistent SCM trajectories, matching the regression baseline without degradation across stabilities.

Finally, we probe the data-driven parameterization under harsh tests where an abrupt transition across stability regimes was imposed. 
From a neutral initialization, the scheme reacts correctly to step forcing in either direction: for neutral$\rightarrow$stable it forms a shallow, surface-based inversion with confined cooling below $\sim 0.3 L_z$, little mixing aloft, and an inertially-modulated low-level jet.
For neutral$\rightarrow$unstable it rapidly homogenizes $\overline{\theta}$, deepens the mixed layer with realistic entrainment warming, and mixes momentum toward vertically uniform winds. 
For direct sign flips, the parameterization captures the expected phenomena, where the temperature adjusts within tens of minutes while winds relax on inertial timescales.
The unsteady$\rightarrow$steady transition produces a residual layer aloft and the emergence of a low-level jet without spurious loft mixing, whereas steady$\rightarrow$unsteady heats the entire vertical, and reduces shear via mixing-down of momentum. 
Key phases (breakup/formation of stability, LLJ height and veering/backing) are captured by the proposed parameterization, where timing and amplitude track the reference LES solution closely, with small, localized errors.

This work provides a first step towards transparent data-driven flux parameterization schemes that offer insight into the ABL physics and emulate LES-derived mean fields faithfully.
At this stage, different questions remain subject to experimentation, namely, would a nonlinear data-driven parameterization perform better than the simpler linear ones tested here?
How does the structure of the linear operator change for heterogeneous surface conditions, and how can we incorporate surface heterogeneity information into the framework? 
How can one account for memory effect in the parameterization for time-dependent forcing problems? 
What modifications would be required for baroclinic conditions or under different Rossby numbers? 
Would it be possible to reverse-engineer the linear operators to express the underlying dynamics analytically?

\clearpage

\appendix 
\section{Derivation of Flux Parameterization}
\label{app:deriv}

We consider the parameterization of turbulent scalar fluxes in the Eddy-Diffusivity Mass-Flux (EDMF) framework, where the vertical flux is expressed as
\begin{equation}
    \overline{w'\theta'} \approx -K \frac{\partial \overline{\theta}}{\partial z} 
    + \underbrace{m_u\big(\theta_u - \overline{\theta}\big)}_{\overline{w'\theta'}_{\text{nonlocal}}}.
\end{equation}

\noindent The nonlocal contribution involves the difference between the updraft temperature $\theta_u$ and the mean value $\overline{\theta}$. 
The evolution of $\theta_u$ with height is modeled as
\begin{equation}
    \frac{\partial \theta_u}{\partial z} = -\epsilon \, \big(\theta_u - \overline{\theta}\big),
\end{equation}
where $\epsilon$ (m$^{-1}$) is the lateral entrainment rate describing the interaction of the updraft with the surrounding environment. 
Note that, for simplicity, we assume that $\epsilon$ is constant in the vertical direction; however, the main conclusions of this Appendix remain unchanged even when $\epsilon$ varies with height.

% ------------------
% ABED: Can probably simplify these VVVVVVVV
\noindent This is a first-order linear ordinary differential equation that can be solved explicitly: 
\begin{equation}
    \frac{\partial \theta_u}{\partial z} + \epsilon \theta_u = \epsilon \overline{\theta}.
\end{equation}

\noindent Introducing $P(z) = \epsilon$ and $Q(z) = \epsilon \overline{\theta}$, the integrating factor is given by
\begin{equation}
    \mu(z) = \exp\left(\int P(z)\,dz\right) = e^{\epsilon z}.
\end{equation}

\noindent Multiplying through by $\mu(z)$ and integrating yields
\begin{equation}
    \frac{d}{dz}\left(e^{\epsilon z}\theta_u\right) = \epsilon e^{\epsilon z}\overline{\theta},
\end{equation}
which gives the solution
\begin{equation}
    \theta_u(z) = e^{-\epsilon z}\left[\int_0^z \epsilon e^{\epsilon s}\overline{\theta}(s)\,ds + C\right],
\end{equation}
with $C$ determined by the boundary condition $\theta_u(0)$. 

% ABED: ^^^^^^^^
% ------------------

\noindent This representation can be written equivalently as a convolution:
\begin{equation}
    \theta_u(z) = r(z) * \overline{\theta}(z),
\end{equation}
where the kernel $r(z)$ is proportional to $e^{-\epsilon z}$ and describes the vertical weighting of the nonlocal contribution. 
This formulation makes explicit that $\theta_u(z)$ depends on the integral influence of $\overline{\theta}$ at all levels below $z$, thereby introducing nonlocal interactions.

\noindent Accordingly, the scalar flux parameterization becomes
\begin{equation}
    w'\theta' = -k \frac{\partial \overline{\theta}}{\partial z} 
    + r(z) * \overline{\theta}(z).
\end{equation}

% \noindent If we denote
% \begin{equation}
%     G(z) = w'\theta' + k \frac{\partial \overline{\theta}}{\partial z},
% \end{equation}
% then $G(z)$, which can be obtained from large-eddy simulations (LES) or measurements, satisfies
% \begin{equation}
%     G(z) = L(z) * \overline{\theta}(z),
% \end{equation}
% where, L is the ``effective'' convolution kernel needed to operate on the mean profile to estimate the corresponding flux.
% By rewriting the flux parameterization equations, a formulation attractive for data-driven studies was reached, allowing for a plethora of novel analyses. 
\noindent Then, given that $k \frac{\partial \overline{\theta}}{\partial z}$ is local, meaning it can also be written as a convolution $q(z) * \overline{\theta}(z)$, we obtain 
\begin{equation}
        \overline{w'\theta'}(z) =L(z)* \overline{\theta}(z) = (q(z) + r(z)) * \overline{\theta}(z),
\end{equation}
where $L=q+r$ is the ``effective'' convolution kernel needed to operate on the mean profile to estimate the corresponding flux.
By rewriting the flux parameterization equations, a formulation attractive for data-driven studies was obtained, allowing for a plethora of novel analyses.

\section{K Profile Parameterization}
\label{app:kpp}

The proposed data-driven methodology was contrasted against the KPP model based on the Troen–Mahrt parameterization \cite{TroenMahrt1986} with a Blackadar mixing length \cite{Blackadar1962}.
The KPP benchmark, vertical transport is parameterized with K-profile eddy coefficients
\begin{equation}
K_{m,h}(z)=\kappa u_* z\,\phi_{m,h}^{-1}\!\left(\frac{z}{L}\right)\,\chi(z),
\end{equation}
where $\kappa=0.4$ is the Von K\`arm\`an coefficient, $\zeta$ the Monin--Obukhov stability parameter, and the used Monin--Obukhov stability functions are
\begin{equation}
\phi_{m,h}(\zeta)=
\begin{cases}
1+5\zeta, & \zeta \ge 0,\\[2pt]
\left(1-16\zeta\right)^{-1/4}, & \zeta < 0 \quad (m),\\[2pt]
\left(1-16\zeta\right)^{-1/2}, & \zeta < 0 \quad (h),
\end{cases}
\qquad \zeta \equiv z/L.
\end{equation}
The boundary layer top is enforced through a quadratic taper
\begin{equation}
\chi(z)=\max\!\left(0,1-\frac{z}{h}\right)^2,
\end{equation}
with prescribed depth $h$, and in unstable conditions, the scalar diffusivity is enhanced by convective scaling,
\begin{equation}
K_h = \kappa u_* z\,\phi_{h}^{-1}\!\left(\frac{z}{L}\right)\,\chi(z) \left[1+\gamma \left(\frac{w_*}{u_*}\right)\max\!\left(0,\frac{z}{h}\left(1-\frac{z}{h}\right)\right)\right],
\end{equation}
with $\gamma=0.7$, $w_*=(B_0h)^{1/3}$, and $B_0=(g/\theta_{\rm ref})\max(\overline{w'\theta'_0},0)$ using $g=9.81~\mathrm{m\,s^{-2}}$ and $\theta_{\rm ref}=300~\mathrm{K}$. Surface momentum and heat fluxes impose Neumann boundary conditions, with stress direction aligned with the instantaneous near-surface wind,
\begin{equation}
(\tau_{u0},\tau_{v0})=-u_*^2\frac{(U,V)}{\sqrt{U^2+V^2}}\bigg|_{z=0},
\end{equation}
and prescribed $\overline{w'\theta'_0}$.
While KPP reproduces large-scale patterns, its errors are substantially larger than those from the operator-based parameterization. 
% This discrepancy reflects the empirical constraints imposed by KPP, which fail to capture the finer-scale dynamics resolved by LES.

\clearpage
%%%%%%%%%%%%%%%%%%%%%%%%%%%%%%%%%%%%%%%%%%%%%%%%%%%%%%%%%%%%%%%%%%%%%
% ACKNOWLEDGMENTS
%%%%%%%%%%%%%%%%%%%%%%%%%%%%%%%%%%%%%%%%%%%%%%%%%%%%%%%%%%%%%%%%%%%%%

%%%%%%%%%%%%%%%%%%%%%%%%%%%%%%%%%%%%%%%%%%%%%%%%%%%%%%%%%%%%%%%%%%%%%%%%%%%%
% DATA SECTION and ACKNOWLEDGMENTS
%%%%%%%%%%%%%%%%%%%%%%%%%%%%%%%%%%%%%%%%%%%%%%%%%%%%%%%%%%%%%%%%%%%%%%%%%%%%

\section*{Open Research Section}
Data accompanying the manuscript are available in a Zenodo repository (10.5281/zenodo.20449690).

\section*{Conflict of Interest declaration}
The authors declare no conflicts of interest for this manuscript.

\section*{Acknowledgments}
This research has been supported by the National Oceanic and Atmospheric Administration (US Department of Commerce grant no. NA23OAR4320198) and Princeton University through the Cooperative Institute for Modeling the Earth System, and by the US National Science Foundation under grant number AGS 2128345. AH was also supported by a postdoctoral fellowship from the Gordon and Betty Moore Foundation.
We would also like to acknowledge high-performance computing support from Princeton University. 
The statements, findings, conclusions, and recommendations are those of the authors and do not necessarily reflect the views of the National Science Foundation, the National Oceanic and Atmospheric Administration or the Department of Commerce.

%%%%%%%%%%%%%%%%%%%%%%%%%%%%%%%%%%%%%%%%%%%%%%%%%%%%%%%%%%%%%%%%%%%%%%%%%%%%
% REFERENCES
%%%%%%%%%%%%%%%%%%%%%%%%%%%%%%%%%%%%%%%%%%%%%%%%%%%%%%%%%%%%%%%%%%%%%%%%%%%%
\bibliographystyle{plainnat}
\bibliography{references}

\end{document}